\documentclass[lettersize, journal, onecolumn]{IEEEtran} 
\usepackage[T1]{fontenc} 
\usepackage{cite}
\usepackage {graphicx}
\usepackage{amssymb}
\usepackage{amsmath}
\usepackage{mathtools}
\usepackage[ruled, linesnumbered]{algorithm2e}
\usepackage[caption=false, font=normalsize, labelfont=sf, textfont=sf]{subfig}
\usepackage{physics}
\usepackage[dvipsnames]{xcolor}
\usepackage{hyperref}
\usepackage{multirow}
\usepackage{stfloats}
\usepackage{mathrsfs}
\usepackage{setspace}

\begin{document}
	\doublespacing
	
\bstctlcite{IEEEexample:BSTcontrol}
	
\title{Quantum Radar for ISAC: Sum-Rate Optimization}

\author{
	Abdulmohsen~Alsaui,
	Octavia~A.~Dobre,
    Neel~Kanth~Kundu,
    Abdulkarim~Hariri,
    and~Hyundong~Shin%

}

\maketitle

\begin{abstract}
	Integrated sensing and communication (ISAC) is emerging as a key enabler for spectrum-efficient and hardware-converged wireless networks. However, classical radar systems within ISAC architectures face fundamental limitations under low signal power and high-noise conditions. This paper proposes a novel framework that embeds quantum illumination radar into a base station to simultaneously support full-duplex classical communication and quantum-enhanced target detection. The resulting integrated quantum sensing and classical communication (IQSCC) system is optimized via a sum-rate maximization formulation subject to radar sensing constraints. The non-convex joint optimization of transmit power and beamforming vectors is tackled using the successive convex approximation technique. Furthermore, we derive performance bounds for classical and quantum radar protocols under the statistical detection theory, highlighting the quantum advantage in low signal-to-interference-plus-noise ratio regimes. Simulation results demonstrate that the proposed IQSCC system achieves a higher communication throughput than the conventional ISAC baseline while satisfying the sensing requirement.
\end{abstract}
 

\IEEEpeerreviewmaketitle

\section{Introduction}
	The rapid proliferation of wireless devices, coupled with the growing demand from spectrum-intensive applications such as autonomous vehicles and the industrial Internet-of-things, necessitates communication systems capable of meeting rising data requirements while also supporting accurate localization capabilities~\cite{trevlakis2023localization, zhou20245g}. In this context, integrated sensing and communication (ISAC) has emerged as a key technology of next-generation wireless networks. ISAC enables the joint use of transmitted signals for both environmental sensing and data communication, resulting in a more efficient spectrum utilization, reduced hardware redundancy, and the ability to support emerging services such as real-time radar imaging with high-throughput data transmission~\cite{lu2024integrated, liu2022integrated}.
	
	Earlier research on ISAC primarily focused on joint waveform and beamforming design guided by classical radar estimation metrics. For example, dual-function radar communication systems have been developed to jointly optimize transmit beamforming for both communication quality and radar sensing accuracy, embedding communication signals into the radar waveform while minimizing target estimation error bounds~\cite{liu2021cramer}. More recent ISAC studies extend this framework to in-band sensing with full-duplex (FD) communication, where simultaneous transmission and reception are supported. These works propose joint beamforming and power control schemes to either maximize the communication sum rate or minimize transmit power, subject to both sensing and communication constraints, thereby improving spectral and energy efficiency~\cite{he2023full}.
	
	Despite these advances, ISAC systems can experience significant degradation in sensing performance under low signal power or high background noise conditions. To maintain acceptable sensing quality, power is often reallocated from communication functionalities, thereby reducing overall throughput or degrading service quality~\cite{lu2024integrated}. Moreover, even without thermal noise, the sensing capabilities of classical radar systems are ultimately limited by quantum noise through the standard quantum limit, which sets a bound on the estimation precision achievable by any classical measurement system at a given signal power~\cite{giovannetti2004quantum}.
	
	To overcome the limitations of classical sensing in low signal power and high noise environments, quantum radar has emerged as a promising paradigm that leverages quantum entanglement to enhance detection performance. The quantum illumination (QI) protocol, introduced by Lloyd in 2008, utilizes entangled photon pairs, where one photon probes the target and the other is retained as a reference, to improve detection probability under high background noise~\cite{lloyd2008enhanced}. Building on this work, a range of specialized receiver architectures have been proposed to approach the theoretical limits of QI, including feed-forward sum-frequency generation, correlation-to-displacement, and other joint measurement schemes~\cite{reichert2023quantum, angeletti2023microwave, barzanjeh2020microwave}. Recent experimental demonstrations in the microwave regime have reported a measurable quantum advantage of around $0.8$~dB over the optimal classical strategy under identical conditions~\cite{assouly2023quantum}. Theoretical analyses indicate that practical joint-measurement receivers may achieve up to $3$~dB gain, while the ultimate performance bound, achievable only with ideal collective measurements, approaches $6$~dB~\cite{guha2009gaussian, tan2008quantum}. Furthermore, recent studies have derived quantum-enhanced radar equations and analyzed detection-range trade-offs, which highlighted how entanglement-obtained gains are affected by photon budget and thermal noise constraints in realistic microwave-band scenarios~\cite{pavan2024range, wei2023evaluating}.
	
	The concept of ISAC has been extended into the quantum domain in several recent studies. In~\cite{yao2025utility}, entanglement is leveraged for simultaneous data transmission and instantaneous target detection, although communication reliability is limited due to encoding information into an extremely weak signal mode. A broader ISAC framework is explored in~\cite{liu2024quantum}, which employs entangled photon pairs for both quantum secure direct communication and remote phase sensing with Heisenberg-limited precision. In~\cite{xu2024integrated}, a field-deployed fiber network demonstrates the coexistence of continuous-variable quantum key distribution and distributed vibration sensing, although quantum resources are used solely for secure key exchange. In~\cite{wang2022joint}, the trade-off between communication rate and sensing accuracy is analyzed under a joint communication and sensing model with unknown channel parameters. This framework is extended in~\cite{munar2024joint_1} to the lossy bosonic channel, a physically motivated model for optical links, where joint reflectivity estimation and data transmission are shown to benefit from quantum measurements in the low-photon regime. Furthermore,~\cite{munar2024joint_2} introduces adversarial considerations by studying joint communication and eavesdropper detection over the same bosonic channel.
	
	However, to the best of our knowledge, no prior work has investigated the use of quantum radar for sensing in an ISAC system while retaining classical communication functionalities. This paper presents a novel framework that integrates QI-based radar sensing into a multi-antenna FD base station (BS), which enables concurrent classical data transmission and quantum-enhanced target detection within a unified architecture. The main contributions of this work are as follows:
	\begin{itemize}
		\item We propose a novel integrated quantum sensing and classical communication (IQSCC) framework that embeds quantum two-mode squeezed vacuum (TMSV) radar within an FD ISAC architecture, enabling simultaneous classical data transmission and quantum-enhanced radar sensing.
		\item We formulate a sum-rate maximization problem that jointly optimizes the transmit power, beamforming vectors, and radar waveform design under a radar signal-to-interference-plus-noise ratio (SINR) constraint. The problem is non-convex, and we address it using successive convex approximation (SCA) techniques.
		\item We derive closed-form expressions for the optimal receive beamformers and utilize an iterative algorithm to solve the joint optimization problem.
		\item We derive the receiver operating characteristic (ROC) performance metrics for classical and quantum radar protocols under Gaussian noise.
	\end{itemize}
	
	The remainder of the paper is organized as follows. Section~\ref{Sec:ISAC_System} introduces the ISAC signal model and formulates the joint optimization problem. Section~\ref{Sec:Radar_Systems} succinctly reviews the detection theory for binary hypothesis testing and derives the ROC parameters for the considered radar protocols. The proposed IQSCC architecture and corresponding simulation results are presented in Section~\ref{Sec:IQSCC_System}. Finally, Section~\ref{Sec:Conclusion} concludes the paper and discusses directions for future research.
	
	\textit{Notation}: Boldface lowercase and uppercase letters represent vectors and matrices, respectively. The superscripts $(\cdot)^T$ and $(\cdot)^H$ denote the transpose and Hermitian transpose, respectively. The $\ell_2$ norm of a vector is denoted by $\lVert \cdot \rVert$, and $|\cdot|$ represents the absolute value of a scalar. The trace of a matrix is written as $\mathrm{Tr}(\cdot)$, and the expectation operator is denoted by $\mathbb{E}[\cdot]$. The identity matrix of size $N \times N$ is written as $\mathbf{I}_N$. $\mathbf{X} \succeq 0$ indicates that the matrix $\mathbf{X}$ is positive semi-definite (i.e., nonnegative eigenvalues). Finally, $\mathbb{C}$ denotes the set of complex-valued numbers.

\section{ISAC System}~\label{Sec:ISAC_System}
	This section discusses the considered model for the ISAC system in addition to the optimization problem formulation.
	\subsection{Model}
		\begin{figure}[!t]
			\centering
			\includegraphics[width=0.35\textwidth]{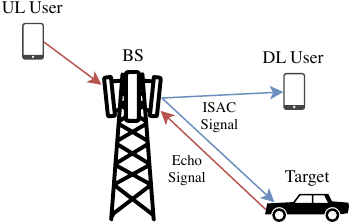}
			\caption{Illustration of the considered ISAC system with a DL user, a UL user, and a monostatic radar target.}
			\label{Fig:ISAC_System_Model}
		\end{figure}
		The considered ISAC system consists of a single BS that simultaneously serves one downlink (DL) user and one uplink (UL) user while performing target sensing, as shown in Fig.~\ref{Fig:ISAC_System_Model}. The DL and UL users are equipped with a single antenna and maintain a line-of-sight link with the BS. The BS is assumed to operate in FD mode for simultaneous transmission and reception with $N_r$ receive and $N_t$ transmit antennas. At the BS, the radar and communication functionalities are jointly performed by the transmitted ISAC signal, expressed as
		\begin{equation}
			\mathbf{x} = \mathbf{v} s + \mathbf{s}_0,
		\end{equation}
		where $\mathbf{v} \in \mathbb{C}^{N_t \times 1}$ is the beamforming vector for the DL user, $s \in \mathbb{C}$ is the unit-power data symbol (i.e., $\mathbb{E} \left[ |s|^2 \right] = 1$), and $\mathbf{s}_0 \in \mathbb{C}^{N_t \times 1}$ is a radar-specific probing signal. The radar signal, $\mathbf{s}_0$, has a covariance matrix defined by $\mathbf{V}_{\hspace{-0.5mm} s} = \mathbb{E} \left[ \mathbf{s}_0 \mathbf{s}_0^H \right] \in \mathbb{C}^{N_t \times N_t}$, which exploits additional spatial degrees of freedom to enhance sensing accuracy. The total transmit power of the BS is upper-limited as follows
		\begin{equation}
			\|\mathbf{v}\|^2 + \mathrm{Tr}(\mathbf{V}_{\hspace{-0.5mm} s}) \le P_b^{\text{max}}.
		\end{equation}
		The UL user transmits a signal $d$ with an average power $p$, which must not exceed the specified power limit
		\begin{equation}
			\mathbb{E}\left[ |d|^2 \right] = p \le P_u^{\text{max}}.
		\end{equation}
		The received signal at the BS is given by
		\begin{equation}
			\mathbf{y}_b = 
			\mathbf{h} d +
			\beta_0 \mathbf{A}(\theta_0) \mathbf{x} +
			\sum_{i=1}^I \beta_i \mathbf{A}(\theta_i) \mathbf{x} +
			\mathbf{H}_{\text{SI}} \mathbf{x} +
			\mathbf{w},
		\end{equation}
		where $\mathbf{h} \in \mathbb{C}^{N_r \times 1}$ denotes the channel between the UL user and BS, $\beta_0$ is the complex amplitude of the target reflection that is located at angle $\theta_0$, $\beta_i$ and $\theta_i$ are the amplitudes and angles of the $I$ environmental interferers, $\mathbf{H}_{\text{SI}} \in \mathbb{C}^{N_r \times N_t}$ models the channel's residual self-interference (SI), and $\mathbf{w}$ is the channel's additive white Gaussian noise (AWGN) with covariance $\sigma_w^2 \mathbf{I}_{N_r}$. The matrix $\mathbf{A}(\theta)$ is defined as follows
		\begin{equation}
			\mathbf{A}(\theta) = \mathbf{a}_r(\theta) \mathbf{a}_t^H(\theta),
		\end{equation}
		where $\mathbf{a}_r(\theta)$ and $\mathbf{a}_t(\theta)$ are the receive and transmit steering vectors, respectively, which are given by
		\begin{equation}
			\mathbf{a}_r(\theta) = \frac{1}{\sqrt{N_r}} [1, e^{j\pi \sin(\theta)}, \dots, e^{j\pi (N_r - 1)\sin(\theta)}]^T,
		\end{equation}
		\begin{equation}
			\mathbf{a}_t(\theta) = \frac{1}{\sqrt{N_t}} [1, e^{j\pi \sin(\theta)}, \dots, e^{j\pi (N_t - 1)\sin(\theta)}]^T.
		\end{equation}
		For the DL user, the received signal can be expressed as follows
		\begin{equation}
			y_d = \mathbf{g}^H \mathbf{v} s + \mathbf{g}^H \mathbf{s}_0 + w,
		\end{equation}
		where $\mathbf{g} \in \mathbb{C}^{N_t \times 1}$ is the channel vector between the DL user and the BS, while $w$ is an AWGN element with variance $\sigma_w^2$. The ISAC system performance depends on the SINRs of each functionality, which for the DL communication is
		\begin{equation}
			\gamma^d = 
			\frac{\mathbf{g}^H \mathbf{V}_{\hspace{-0.5mm} c} \mathbf{g}}
			{\mathbf{g}^H \mathbf{V}_{\hspace{-0.5mm} s} \mathbf{g} + \sigma_w^2},
		\end{equation}
		where $\mathbf{V}_{\hspace{-0.5mm} c} = \mathbf{v} \mathbf{v}^H \in \mathbb{C}^{N_t \times N_t}$. The covariance of the transmitted ISAC signal is defined as $\mathbf{V}_{\hspace{-0.5mm} t} = \mathbf{V}_{\hspace{-0.5mm} c} + \mathbf{V}_{\hspace{-0.5mm} s}$. Thus, the radar SINR after applying a receive beamformer, $\mathbf{u}$, is~\cite{he2023full}
		\begin{equation}
			\label{Eq:Radar_SINR_1}
			\gamma^s = 
			\frac{|\beta_0|^2 \mathbf{u}^H \mathbf{A}(\theta_0) \mathbf{V}_{\hspace{-0.5mm} t} \mathbf{A}(\theta_0)^H \mathbf{u}}
			{\mathbf{u}^H ( p \mathbf{h}\mathbf{h}^H + \mathbf{B} \mathbf{V}_{\hspace{-0.5mm} t} \mathbf{B}^H + \sigma_w^2 \mathbf{I}_{N_r} ) \mathbf{u}},
		\end{equation}
		where $\mathbf{B} = \sum_{i=1}^I \beta_i \mathbf{A}(\theta_i) + \mathbf{H}_{\text{SI}}$. Similarly, the SINR of the UL user is given by
		\begin{equation}
			\label{Eq:Uplink_SINR}
			\gamma^u = 
			\frac{p\, \mathbf{m}^H \mathbf{h} \mathbf{h}^H \mathbf{m}}
			{\mathbf{m}^H ( \mathbf{C} \mathbf{V}_{\hspace{-0.5mm} t} \mathbf{C}^H + \sigma_w^2 \mathbf{I}_{N_r} ) \mathbf{m}},
		\end{equation}
		where $\mathbf{m}$ is a receive beamformer and $\mathbf{C} = \mathbf{B} + \beta_0 \mathbf{A}(\theta_0)$.
		
	\subsection{Problem Formulation}
		The objective is to maximize the system's overall sum rate while ensuring that the minimum required radar SINR is achieved. This is done by jointly designing the UL user transmit power, $p$, the BS transmit beamformers, $\mathbf{v}$ and $\mathbf{V}_{\hspace{-0.5mm} s}$, and the BS receive beamformers, $\mathbf{u}$ and $\mathbf{m}$. The optimal receive beamformers, derived in Subsection~\ref{App:Receive_Beamformers} of the Appendix, are given by
		\begin{equation}
			\label{Eq:Receive_Beamformer_u}
			\mathbf{u}^* = (p\mathbf{h}\mathbf{h}^H + \mathbf{B} \mathbf{V}_{\hspace{-0.5mm} t} \mathbf{B}^H + \sigma_w^2 \mathbf{I}_{N_r})^{-1} \mathbf{a}_r(\theta_0),
		\end{equation}
		\begin{equation}
			\label{Eq:Receive_Beamformer_w}
			\mathbf{m}^* = (\mathbf{C} \mathbf{V}_{\hspace{-0.5mm} t} \mathbf{C}^H + \sigma_w^2 \mathbf{I}_{N_r})^{-1} \mathbf{h},
		\end{equation}
		which when plugged into the respective SINR expressions and simplifying result in
		\begin{equation}
			\label{Eq:Radar_SINR_2}
			\bar{\gamma}^s = |\beta_0|^2 \mathbf{a}_t^H(\theta_0) \mathbf{V}_{\hspace{-0.5mm} t} \mathbf{a}_t(\theta_0) \mathbf{a}_r^H(\theta_0) \mathbf{\Psi}^{-1} \mathbf{a}_r(\theta_0),
		\end{equation}
		\begin{equation}
			\bar{\gamma}^u = p \mathbf{h}^H \mathbf{\Phi}^{-1} \mathbf{h},
		\end{equation}
		where $\mathbf{\Phi} = \mathbf{C} \mathbf{V}_{\hspace{-0.5mm} t} \mathbf{C}^H + \sigma_w^2 \mathbf{I}_{N_r}$ and $\mathbf{\Psi} = p\mathbf{h}\mathbf{h}^H + \mathbf{B} \mathbf{V}_{\hspace{-0.5mm} t} \mathbf{B}^H + \sigma_w^2 \mathbf{I}_{N_r} \in \mathbb{C}^{N_r \times N_r}$~\cite{he2023full}. Thus, the optimization problem can be expressed as
		\begin{equation}
			\label{Eq:Optimization_Problem_1}
			\begin{split}
				\underset{\mathbf{v},\, \mathbf{V}_{\hspace{-0.5mm} s} \succcurlyeq \mathbf{0},\, p \ge 0 }{\text{maximize}} \quad & B\left[ \log_2\left(1 + \bar{\gamma}^u\right) + \log_2 \left(1 + \gamma^d\right) \right] \\
				\text{subject to} \quad & \lVert \mathbf{v} \rVert^2 + \text{Tr}\left(\mathbf{V}_{\hspace{-0.5mm} s}\right) \le P_b^{\text{max}}, \; p \le P_u^{\text{max}}, \\
				& \bar{\gamma}^s \ge \rho^s,
			\end{split}
		\end{equation}
		where $B$ is the system bandwidth and $\rho^s$ is the minimum required radar SINR to achieve a certain desired sensing performance. The objective function and radar SINR constraint of~(\ref{Eq:Optimization_Problem_1}) are nonconcave and nonconvex, respectively. To handle the nonconcavity, define an auxiliary optimization variable $0 \le u \le \bar{\gamma}^u$ and replace $\bar{\gamma}^u$ in~(\ref{Eq:Optimization_Problem_1}) with it. For the second objective function term, reexpress it as follows
		\begin{equation}
			\label{Eq:DL_Sum-Rate_1}
			\begin{split}
				\log_2\left(1 + \gamma^d\right) &= \log_2\left(1 + \frac{\mathbf{g}^H \mathbf{V}_{\hspace{-0.5mm} c} \mathbf{g}}{\mathbf{g}^H \mathbf{V}_{\hspace{-0.5mm} s} \mathbf{g} + \sigma_w^2}\right) \\[1mm]
				&= \log_2\left(\mathbf{g}^H \mathbf{V}_{\hspace{-0.5mm} t} \mathbf{g} + \sigma_w^2\right) \\
				& \hspace{3mm} - \log_2\left(\mathbf{g}^H \mathbf{V}_{\hspace{-0.5mm} s} \mathbf{g} + \sigma_w^2\right),
			\end{split}
		\end{equation}
		where both terms are concave with respect to $\mathbf{V}_{\hspace{-0.5mm} s}$ and $\mathbf{V}_{\hspace{-0.5mm} c}$. For a concave function, $f(x)$, its first-order expansion around $x_0$ satisfies the following inequality
		\begin{equation}
			f(x) \le f(x_0) +\left[f'(x_0)\right]^H (x - x_0).
		\end{equation}
		Thus, using the SCA technique to linearize the second term gives
		\begin{equation}
			\begin{split}
				\log_2\left(\mathbf{g}^H \mathbf{V}_{\hspace{-0.5mm} s} \mathbf{g} + \sigma_w^2\right) & \le \log_2\left( \mathbf{g}^H \mathbf{V}_{\hspace{-0.5mm} s}^{(j-1)} \mathbf{g} + \sigma_w^2 \right) \\[4mm]
				& \quad + \frac{\mathbf{g}^H \left(\mathbf{V}_{\hspace{-0.5mm} s} - \mathbf{V}_{\hspace{-0.5mm} s}^{(j-1)}\right) \mathbf{g}}{\left( \mathbf{g}^H \mathbf{V}_{\hspace{-0.5mm} s}^{(j-1)} \mathbf{g} + \sigma_w^2 \right) \ln(2)},
			\end{split}
		\end{equation}
		where $(j - 1)$ is the iteration's index. Substituting this into~(\ref{Eq:DL_Sum-Rate_1}) yields a concave lower bound for the DL sum-rate term
		\begin{equation}
			\begin{split}
				\log_2\left(1 + \gamma^d\right) & \ge \log_2\left(\mathbf{g}^H \mathbf{V}_{\hspace{-0.5mm} t} \mathbf{g} + \sigma_w^2\right) \\[4mm]
				& \hspace{3mm} - \log_2\left( \mathbf{g}^H \mathbf{V}_{\hspace{-0.5mm} s}^{(j-1)} \mathbf{g} + \sigma_w^2 \right) \\[4mm]
				& \hspace{3mm} - \frac{\mathbf{g}^H \left(\mathbf{V}_{\hspace{-0.5mm} s} - \mathbf{V}_{\hspace{-0.5mm} s}^{(j-1)}\right) \mathbf{g}}{\left( \mathbf{g}^H \mathbf{V}_{\hspace{-0.5mm} s}^{(j-1)} \mathbf{g} + \sigma_w^2 \right) \ln(2)}.
			\end{split}
		\end{equation}
		Next, to handle the nonconvexity of the radar SINR constraint, it is first rewritten as
		\begin{equation}
			\label{Eq:Classical_Radar_SINR_Constraint}
			\mathbf{a}_r^H(\theta_0)\mathbf{\Psi}^{-1} \mathbf{a}_r(\theta_0) \ge \frac{\rho^s}{|\beta_0|^2} \left[\mathbf{a}_t^H(\theta_0) \mathbf{V}_{\hspace{-0.5mm} t} \mathbf{a}_t(\theta_0) \right]^{-1},
		\end{equation}
		where the fact that $\mathbf{V}_{\hspace{-0.5mm} t} \succeq 0$ and $\mathbf{a}_t^H(\theta_0)\mathbf{V}_{\hspace{-0.5mm} t}\mathbf{a}_t(\theta_0) > 0$ have been used. The left-hand side is convex with respect to $\mathbf{\Psi}$ as well as $p$ and $\mathbf{V}_{\hspace{-0.5mm} t}$, since $\mathbf{\Psi}$ is an affine function of them~\cite{boyd2004convex}. Similarly, the right-hand side is convex with respect to $\mathbf{V}_{\hspace{-0.5mm} t}$. This difference-of-convex form can be handled by linearizing the left-hand side using its first-order Taylor expansion. Using the identity for the derivative of a quadratic form in a matrix inverse~\cite{hjorungnes2011complex, magnus2019matrix}, the expansion yields
		\begin{equation}
			\begin{split}
					\mathbf{a}_r^H(\theta_0)\mathbf{\Psi}^{-1} \mathbf{a}_r(\theta_0) &\ge \mathbf{a}_r^H(\theta_0)\left(\mathbf{\Psi}^{(j-1)}\right)^{-1} \mathbf{a}_r(\theta_0) \\
					& -\mathbf{a}_r^H(\theta_0) \left(\mathbf{\Psi}^{(j-1)} \right)^{-1} \hspace{-1mm} \\
					& \hspace{3mm} \left(\mathbf{\Psi} - \mathbf{\Psi}^{(j-1)}\right) \hspace{-1mm} \left(\mathbf{\Psi}^{(j-1)} \right)^{-1} \hspace{-3.5mm} \mathbf{a}_r(\theta_0) .
			\end{split}
		\end{equation}
		Performing a similar expansion on the UL SINR constraint gives
		\begin{equation}
			\begin{split}
				\frac{u}{p} & \le \mathbf{h}^H\left(\mathbf{\Phi}^{(j-1)}\right)^{-1} \mathbf{h}  \\ 
				& \hspace{3.5mm} - \mathbf{h}^H \left(\mathbf{\Phi}^{(j-1)} \right)^{-1} \left(\mathbf{\Phi}
				- \mathbf{\Phi}^{(j-1)}\right) \left(\mathbf{\Phi}^{(j-1)} \right)^{-1}\mathbf{h}.
			\end{split}
		\end{equation}
		To convexify the fraction, an auxiliary real variable, $x$, is introduced to break it into two constraints
		\begin{equation}
			\label{Eq:Uplink_SINR_Constraint}
			\hspace{-0.5mm}
			\begin{cases}
				\frac{x^2}{p} \le \mathbf{h}^H\left(\mathbf{\Phi}^{(j-1)}\right)^{-1} \mathbf{h}  \\
				\hspace{7.5mm} -\mathbf{h}^H \left(\mathbf{\Phi}^{(j-1)} \right)^{-1} \hspace{-1mm} \left(\mathbf{\Phi} - \mathbf{\Phi}^{(j-1)}\right) \hspace{-1mm} \left(\mathbf{\Phi}^{(j-1)} \right)^{-1} \mathbf{h}, \\
				u \le x^2,
			\end{cases}
			\hspace{-1.5mm}
		\end{equation}
		where the first constraint is convex since $p > 0$~\cite{boyd2004convex}. The second constraint can be linearized by a first-order Taylor approximation
		\begin{equation}
			u \le \left(x^{(j-1)}\right)^2 + 2x^{(j-1)}\left(x - x^{(j-1)} \right).
		\end{equation}
		Putting everything together, the overall convex optimization problem becomes
		\begin{equation}
			\label{Eq:Optimization_Problem_2}
			\begin{split}
					\underset{\substack{\\ p \ge 0,\, u\ge0,\, x\ge0,\, \\[1mm] \{\mathbf{V}_{\hspace{-0.5mm} s}, \mathbf{V}_{\hspace{-0.5mm} c}\} \succcurlyeq \mathbf{0}}}{\text{maximize}} \quad & B \log_2\left(1 + u \right) + B \log_2\left(\mathbf{g}^H \mathbf{V}_{\hspace{-0.5mm} t} \mathbf{g} + \sigma_w^2\right) \hspace{-2mm} \\[-5mm]
					& - B \log_2\left( \mathbf{g}^H \mathbf{V}_{\hspace{-0.5mm} s}^{(j-1)} \mathbf{g} + \sigma_w^2 \right) \\[1mm]
					& - B \frac{\mathbf{g}^H \left(\mathbf{V}_{\hspace{-0.5mm} s} - \mathbf{V}_{\hspace{-0.5mm} s}^{(j-1)} \right) \mathbf{g}}{\left( \mathbf{g}^H \mathbf{V}_{\hspace{-0.5mm} s}^{(j-1)} \mathbf{g} + \sigma_w^2 \right) \ln(2)} \\[3mm]
					\text{subject to} \quad & \text{Tr}\left(\mathbf{V}_{\hspace{-0.5mm} t}\right) \le P_b^{\text{max}}, \; p \le P_u^{\text{max}}, \; (\ref{Eq:Classical_Radar_SINR_Constraint}), \; (\ref{Eq:Uplink_SINR_Constraint}), \hspace{-2mm}
			\end{split}
		\end{equation}
		which is solved using Algorithm~\ref{Alg:SCA}, where the optimization variables are initialized to satisfy basic domain feasibility, that is, $\mathbf{V}_s^{(0)}, \mathbf{V}_c^{(0)} \succeq 0$ and $p^{(0)}, x^{(0)} \ge 0$.
		
		\begin{algorithm}[!t]
	\caption{Sum-Rate Maximization Algorithm}
	\label{Alg:SCA}
	\setstretch{1.5}
	Initialize $\mathbf{V}_{\hspace{-0.5mm} s}^{(0)} = 0.5 \, P_b^{\max} \, \mathbf{a}_t(\theta_0) \mathbf{a}_t^{H}(\theta_0)$, $\mathbf{V}_{\hspace{-0.5mm} c}^{(0)} = 0.5 \, P_b^{\max} \, \mathbf{g} \mathbf{g}^{H}$, $p^{(0)} = P_u^{\max}$, and $x^{(0)} = \left[ p^{(0)} \, \mathbf{h}^{H}\boldsymbol{\Phi}^{-1} \left(\mathbf{V}_{\hspace{-0.5mm} s}^{(0)} + \mathbf{V}_{\hspace{-0.5mm} c}^{(0)} \right) \mathbf{h} \right]^{1/2}$ for $j=0$\;
	\setstretch{1}
	\While{\textnormal{not converged}}{
		$j \leftarrow j + 1$\;
		Solve~(\ref{Eq:Optimization_Problem_2}) and update $\mathbf{V}_{\hspace{-0.5mm} s}^{(j)}$, $\mathbf{V}_{\hspace{-0.5mm} c}^{(j)}$, $p^{(j)}$, and $x^{(j)}$\;
	}
\end{algorithm}
\section{Radar Systems}~\label{Sec:Radar_Systems}
	This section introduces the fundamentals of binary hypothesis detection theory as applied in radar systems. It then discusses the classical radar protocols commonly used as benchmarks for quantum radar implementations, followed by a performance assessment of the considered quantum and classical radar protocols.
	\subsection{Detection Theory}
		In radar signal processing, the detection task is formulated as a binary hypothesis testing problem
		\begin{equation}
			\begin{split}
				& \mathcal{H}_0: \text{ Target Absent } (\eta^s = 0), \\[2mm]
				& \mathcal{H}_1: \text{ Target Present } (0 < \eta^s < 1),
			\end{split}
		\end{equation}
		where $\eta^s$ denotes the overall channel transmissivity, which encompasses both the propagation loss and the target reflection coefficient, and is assumed to be known and fixed under $\mathcal{H}_1$. A suitable test statistic, typically representing the average output of a detector over $K$ measurements, is used to distinguish between the two hypotheses and is defined as
		\begin{equation}
			\label{Eq:Test_Statistic}
			\mathcal{T} = \frac{1}{K} \sum_{k=1}^K \mathcal{T}^{[k]} \sim
			\begin{cases}
				\mathcal{N} \left( \mu_0, \sigma^2_0 \right), & \text{under } \mathcal{H}_0, \\
				\mathcal{N} \left( \mu_1, \sigma^2_1 \right), & \text{under } \mathcal{H}_1.
			\end{cases}
		\end{equation}
		This formulation, which assumes Gaussian-distributed test statistics, holds under the central limit theorem for a large number of independent and identically distributed measurements. The mean and variance of the test statistic under each hypothesis are given by
		\begin{equation}
			\mu_0 = \mathbb{E} \left[ \mathcal{T}^{[k]} \middle| \mathcal{H}_0 \right] = 0,
		\end{equation}
		\begin{equation}
			\mu_1 = \mathbb{E} \left[ \mathcal{T}^{[k]} \middle| \mathcal{H}_1 \right],
		\end{equation}
		\begin{equation}
			\sigma_0^2 = \frac{\text{Var} \left[ \mathcal{T}^{[k]} \middle| \mathcal{H}_0 \right] }{K} = \frac{\sigma'^2_0}{K},
		\end{equation}
		\begin{equation}
			\sigma_1^2 = \frac{\text{Var} \left[ \mathcal{T}^{[k]} \middle| \mathcal{H}_1 \right]}{K} = \frac{\sigma'^2_1}{K}.
		\end{equation}
		The two key performance metrics in radar detection are the probability of detection, $P_d$, and the false alarm (FA) probability, $P_f$. Specifically, $P_d$ denotes the probability of correctly detecting a target when it is present, whereas $P_f$ refers to the probability of incorrectly declaring a target when none is present. The ROC provides a comprehensive measure of radar system performance by capturing the trade-off between $P_d$ and $P_f$
		\begin{equation}
			\label{Eq:ROC}
			P_d\left( P_f \right) = Q\left(\mathcal{A}_1 Q^{-1}\left(P_f\right) - \mathcal{A}_2 \right),
		\end{equation}
		where $Q(\cdot)$ is the complementary cumulative distribution function of the standard normal distribution and the auxiliary parameters are introduced to ease the radar protocols analysis~\cite{bischeltsrieder2024engineering}
		\begin{equation}
			\label{Eq:Radar_Parameters}
			\mathcal{A}_1 = \frac{\sigma'_0}{\sigma'_1} = \frac{\sigma_0}{\sigma_1}, \qquad \mathcal{A}_2 = \frac{\mu_1 \sqrt{K}}{\sigma'_1} = \frac{\mu_1}{\sigma_1}.
		\end{equation}
		The derivation of~(\ref{Eq:ROC}) is shown in Subsection~\ref{App:ROC} of the Appendix.
		
	\subsection{Protocols Modeling}
\begin{figure}[!t]
			\centering
			\includegraphics[width=0.48\textwidth]{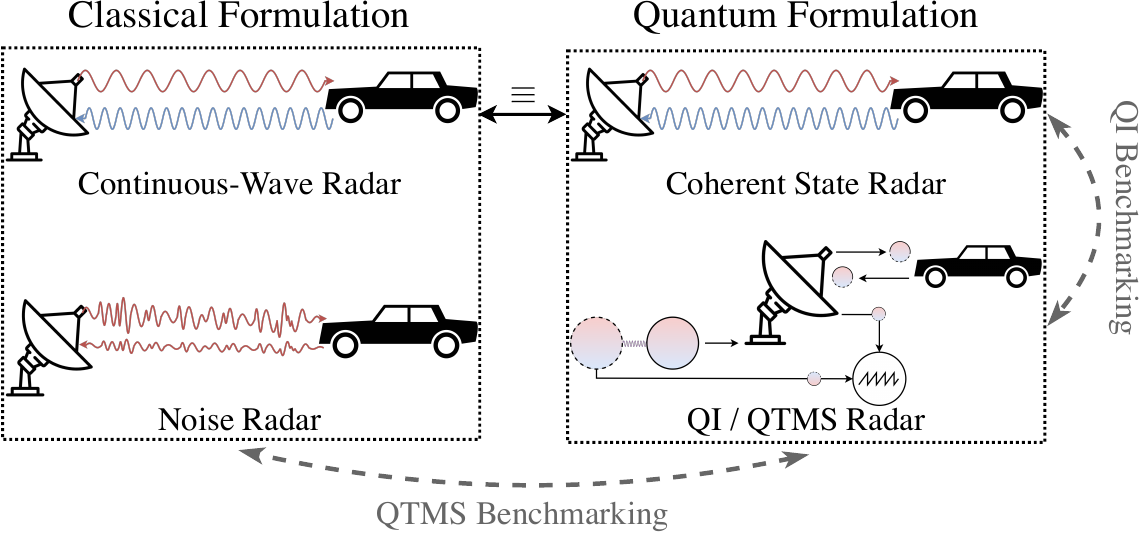}
			\caption{The benchmarking radars used to evaluate the performance of each quantum radar.}
			\label{Fig:Radar_Protocols}
		\end{figure}

		Quantum radar aims to enhance target detection by using quantum entanglement as an additional resource, enabling the identification of targets even in environments dominated by thermal noise. The prominent protocol in this field is QI, which employs entangled photon pairs generated in a TMSV state. One photon, referred to as the signal, is transmitted toward the target region, while its entangled partner, the idler, is stored locally and used in a joint measurement with the returned signal~\cite{lloyd2008enhanced}. A commonly studied special case of QI is the quantum two-mode squeezing (QTMS) protocol, which simplifies implementation by measuring the signal and idler separately rather than a quantum joint measurement~\cite{luong2019receiver}. Each protocol is benchmarked against a classical counterpart, as shown in Fig.~\ref{Fig:Radar_Protocols}. The quantum TMSV radar is compared to the quantum formulation of continuous-wave (CW) radar, known as coherent-state radar, which uses a single-frequency sinusoidal signal to probe the target~\cite{bischeltsrieder2024engineering}. In contrast, QTMS is evaluated relative to the classical noise radar, which utilizes correlated random noise waveforms that are spread across a wide bandwidth, making it resistant to interference and interception~\cite{luong2019receiver}. In this work, we consider the quantum TMSV protocol incorporating a correlation receiver, offering a theoretical $3$~dB advantage in the error probability exponent under conditions of low signal power and high background noise~\cite{bischeltsrieder2024engineering}. Moreover, it demonstrates resilience to environmental decoherence and presents the most promising path toward realizing a genuine quantum enhancement in radar detection~\cite{tan2008quantum}. In the following section, we derive the relevant parameters needed to evaluate~(\ref{Eq:ROC}) for each considered radar protocol.
		
		\subsubsection{Classical Continuous-Wave (CW) Radar}
			The classical analog to the quantum TMSV radar is the CW radar. It transmits a single unmodulated tone of constant frequency and amplitude over time. While it does not provide range information, it is capable of detecting the presence and velocity of a moving target through the Doppler shift induced by relative motion~\cite{richards2010principles}. The received signal at time $\tau$ is modeled as
			\begin{equation}
				\label{Eq:CW_Radar_Model}
				s_r(\tau) = \sqrt{\eta^s} s_t + w(\tau),
			\end{equation}
			where $s_t(\tau) = x_t \cos(2 \pi f \tau)$ is a deterministic tone of frequency $f$ and amplitude $x_t$ directed at the target and $w(\tau)$ is the AWGN sample. The noise is decomposed into its quadrature components as
			\begin{equation}
				\begin{split}
					w(\tau) &=  w_x(\tau) \cos(2 \pi f \tau) + w_p(\tau) \sin(2 \pi f \tau) \\
					& \sim \mathcal{N}\left(0, \sigma_w^2 \right),
				\end{split}
			\end{equation}
			where $ w_x(\tau)$ and $w_p(\tau)$ are independent zero-mean Gaussian processes with variance $\sigma_w^2$. By demodulating the received signal and sampling the in-phase (cosine) component, the resulting signal at sampling instant $\tau_s k$ is
			\begin{equation}
				\label{Eq:CW_Received_x_Quadrature}
				s_{r, x}(\tau_s k) = \sqrt{\eta^s} x_t +  w_x(\tau_s k) \sim \mathcal{N} \left( \sqrt{\eta^s} \, x_t, \sigma_w^2 \right),
			\end{equation}
			where $k$ is the sample index and $\tau_s$ is the sampling interval. The mean signal power, in the absence of noise, is computed as
			\begin{equation}
				\lim_{\tau' \to \infty} \frac{1}{\tau'} \int_{- \tau' / 2}^{\tau' / 2} \mathbb{E}\left[ s_r^2(\tau) | \sigma_w^2 = 0 \right] d\tau = \frac{\eta^s x_t^2}{2},
			\end{equation}
			which leads to the SINR expression for CW radar
			\begin{equation}
				\label{Eq:CW_SINR}
				\gamma^{\text{CW}} = \frac{\eta^s x_t^2}{2 \sigma_w^2}.
			\end{equation}
			Substituting the SINR from~(\ref{Eq:CW_SINR}) along with the variance and mean from~(\ref{Eq:CW_Received_x_Quadrature}) into~(\ref{Eq:Radar_Parameters}), the radar performance parameters for the CW protocol are identified as
			\begin{equation}
				\label{Eq:CW_Radar_Parameters}
				\mathcal{A}_1^{\text{CW}} = 1, \qquad \mathcal{A}_2^{\text{CW}} =  \sqrt{2 \gamma K}.
			\end{equation}
			
\begin{figure}[!t]
				\centering
				\includegraphics[width=0.48\textwidth]{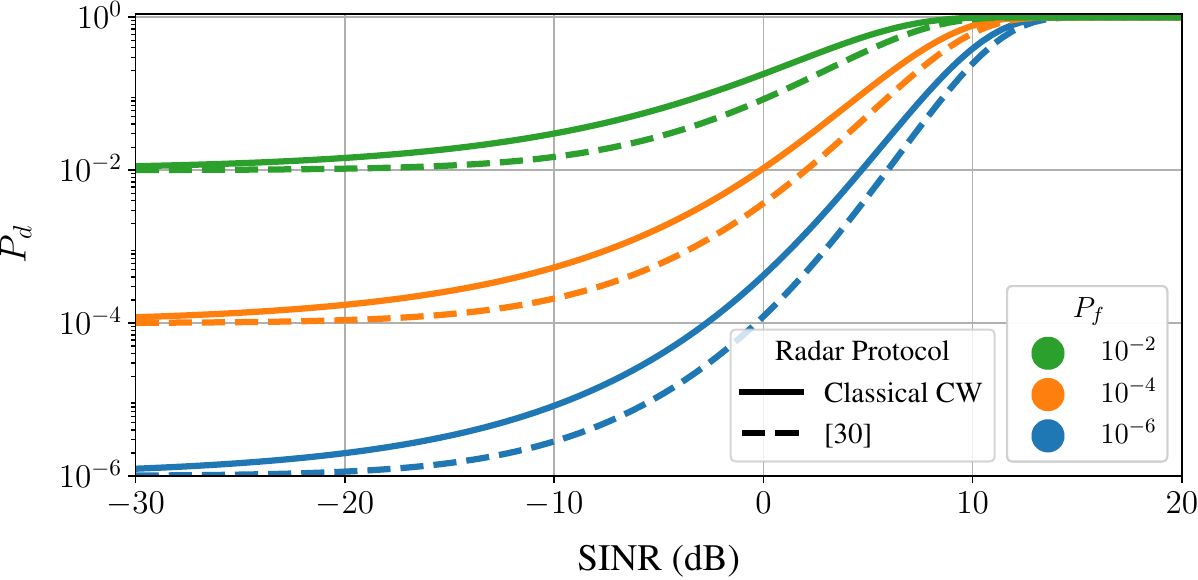}
				\caption{Detection probability versus SINR for the derived classical CW and reference radar models.}
				\label{Fig:Pd_Vs_SINR_Classical_Protocols}
			\end{figure}

			\noindent To validate the derived performance parameters of the classical CW radar benchmark, we compare them against standard expressions reported in~\cite{kay1993fundamentals} and~\cite{de2008code}. Specifically, the ROCs are given by
			\begin{equation}
				P_d \left(P_f\right)^{\text{\cite{kay1993fundamentals}}} = Q\left(Q^{-1} \left( P_f \right) - \sqrt{K \gamma} \right),
			\end{equation}
			\begin{equation}
				P_d \left(P_f\right)^{\text{\cite{de2008code}}} = Q_1\left(\sqrt{2 \gamma}, \sqrt{-2 \ln{\left( P_f \right)}} \right),
			\end{equation}
			where $Q_n(\alpha, \beta)$ denotes the $n^{\text{th}}$-order Marcum Q-function, defined as
			\begin{equation}
				Q_n(\alpha, \beta) = \frac{1}{\alpha^{n - 1}} \int_{\beta}^{\infty} x^n e^{-\frac{x^2 + \alpha^2}{2}} I_n(\alpha x) dx,
			\end{equation}
			and $I_k(\cdot)$ is the modified Bessel function of the first kind of order $k$. Note that the expression from~\cite{kay1993fundamentals} differs from the derived CW model in~(\ref{Eq:CW_Radar_Parameters}) by a factor of two in the SINR term. This discrepancy arises because~\cite{kay1993fundamentals} defines the noise variance over a single quadrature (the in-phase component), whereas the CW derivation models the complex signal with both in-phase and quadrature components. Since noise is independent in each quadrature, the total noise variance is doubled. Once this difference is accounted for, the two models are physically equivalent. Fig.~\ref{Fig:Pd_Vs_SINR_Classical_Protocols} plots the detection probability as a function of SINR for several FA probabilities, $P_f$, comparing the derived CW radar model with that of~\cite{de2008code}. As expected, the detection probability is lower-bounded by $P_f$. The consistently better performance of the CW model is expected, as it assumes coherent detection, whereas~\cite{de2008code} considers non-coherent detection.
			
		\subsubsection{Classical Coherent State (CS) Radar}
			The quantum formulation of the classical CW radar provides a reliable benchmark for the quantum TMSV radar. In this formulation, the quantum channel behaves as a lossy beamsplitter, conserving the total energy by partitioning it between signal and noise components. Quantum optical modes are described using the annihilation operator, $\hat{a}$, which defines the coherent state (CS), $\ket{\alpha}$, as its eigenstate
			\begin{equation}
				\hat{a} \ket{\alpha} = \alpha \ket{\alpha},
			\end{equation}
			where $\alpha \in \mathbb{C}$ is the complex-valued eigenvalue. In the number (Fock) basis, this CS is represented by~\cite{weedbrook2012gaussian}
			\begin{equation}
				\ket{\alpha} = e^{-|\alpha|^2/2} \sum_{n = 0}^{\infty} \frac{\alpha^n}{\sqrt{n!}} \ket{n},
			\end{equation}
			where $n$ denotes the photon number. The received field operator for the $k^{\text{th}}$ temporal mode is modeled as a linear combination of the transmitted signal, $\hat{a}_t^{[k]}$, and thermal noise, $\hat{a}_n^{[k]}$~\cite{bischeltsrieder2024engineering}
			\begin{equation}
				\label{Eq:Quantum_Channel_Received_Signal}
				\hat{a}_r^{[k]} = \sqrt{\eta^s} \, \hat{a}_t^{[k]} + \sqrt{1 - \eta^s} \, \hat{a}_n^{[k]}.
			\end{equation}
			
			\begin{figure}[!t]
				\centering
				\includegraphics[width=0.48\textwidth]{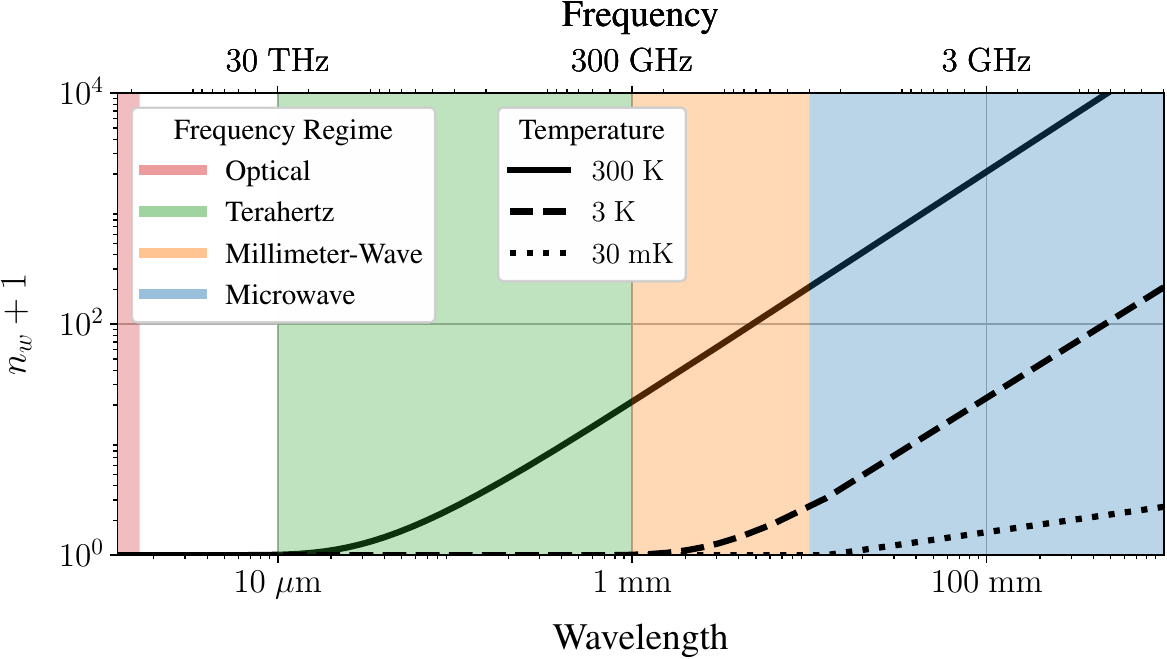}
				\caption{Thermal photon count as a function of frequency and background temperature.}
				\label{Fig:NoisePhotons_Vs_Wavelength}
			\end{figure}
			
			\noindent This is analogous to a beamsplitter transmitting a fraction $\eta^s$ of the signal and injecting thermal noise in the remaining fraction. The receiver performs a homodyne measurement of the in-phase quadrature component, $\hat{x}^{[k]}$. As derived in Subsection~\ref{App:CS_Radar_Parameters} of the Appendix, the radar performance parameters for the CS protocol are given by
			\begin{equation}
				\label{Eq:CS_Radar_Parameters}
				\mathcal{A}_1^{\text{CS}} = 1, \qquad \mathcal{A}_2^{\text{CS}} = 2 \sqrt{\frac{\gamma n_{w} K}{2 n_{w} + 1}},
			\end{equation}
			where $\gamma = \eta^s n_s / n_w$ denotes the SINR and $n_{w}$ is the thermal photon number as determined by the Bose--Einstein distribution in~(\ref{Eq:Bose-Einstein_Distribution}). Fig.~\ref{Fig:NoisePhotons_Vs_Wavelength} shows how the thermal photon count varies with frequency and background temperature. The solid curve, which corresponds to a typical background temperature, is the most relevant in practical settings. In the optical frequency range, the thermal noise is nearly negligible. This is one of the primary reasons why quantum communication systems are often implemented in the optical domain~\cite{zhang2023millimeter, alsaui2024machine}. Conversely, the QI protocols offer advantages primarily in high-noise environments. For this reason, most implementations focus on the microwave regime~\cite{assouly2023quantum, pavan2024range, bischeltsrieder2024engineering, sorelli2021detecting, shapiro2020quantum, wei2023evaluating, luong2019receiver}. Additionally, entangled photon pairs can be readily generated in the microwave range using existing hardware, unlike in the millimeter-wave and terahertz bands~\cite{wei2023evaluating}.
			
			\begin{figure}[!t]
				\centering
				\includegraphics[width=0.48\textwidth]{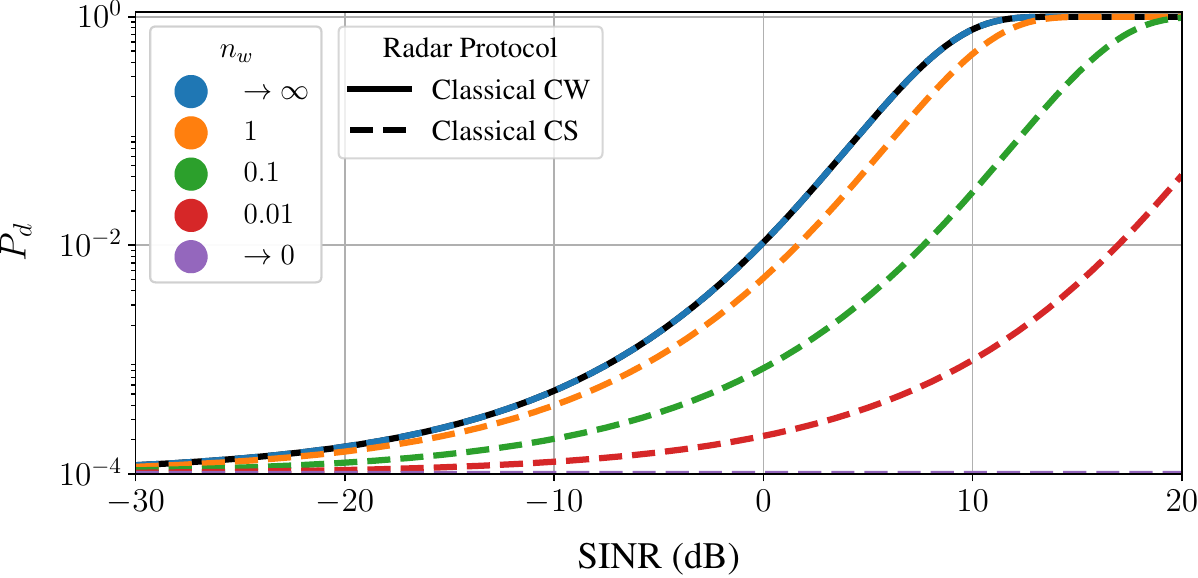}
				\caption{Detection probability as a function of the SINR for the classical CW and CS radar models for different thermal noise photon count numbers, with $P_f = 10^{-4}$.}
				\label{Fig:Pd_Vs_SINR_Classical_Quantum_Formulation}
			\end{figure}
			
			The performance of the CS radar model is compared with that of the classical CW protocol in Fig.~\ref{Fig:Pd_Vs_SINR_Classical_Quantum_Formulation} for various thermal noise levels, with the FA probability fixed at $P_f = 10^{-4}$. As expected, the CS radar's performance converges to that of the classical CW radar as the number of thermal noise photons increases
			\begin{equation}
				\lim_{n_{w} \rightarrow \infty} \mathcal{A}_2^{\text{CS}} = \mathcal{A}_2^{\text{CW}}.
			\end{equation}
			However, as the thermal noise decreases, the detection probability of the CS radar begins to degrade. This degradation occurs because, for a fixed SINR, the signal power also becomes extremely small when $n_{w}$ is low, placing the system in the quantum regime. In this regime, the assumptions underpinning the classical CW formulation no longer hold. This motivates the use of the CS radar as a more physically accurate benchmark for evaluating the quantum TMSV radar performance, which typically operates with weak signal powers.
		\subsubsection{Quantum TMSV Radar}
			The quantum TMSV radar employs entangled photon pairs, expressed as
			\begin{equation}
				\ket{\psi} = \sum_{n' = 0}^{\infty} \sqrt{\frac{n_q^{n'}}{\left(1 + n_q\right)^{n' + 1}}} \ket{n'}_t \ket{n'}_i,
			\end{equation}
			where $n_q$ denotes the average photon number in each mode. The subscripts $t$ and $i$ refer to the transmitted signal and the retained idler, respectively. The associated radar parameters are derived in Subsection~\ref{App:TMSV_Radar_Parameters} of the Appendix, and are given by 
			\begin{equation}
				\label{Eq:TMSV_Radar_Parameter_1}
				\mathcal{A}_1^{\text{TMSV}} = \sqrt{1 + \frac{ 4 \gamma n_{w} + 3 \eta^s }{2 n_{w} + 1+ \frac{\eta^s}{\gamma} \left( 1 + \frac{1}{n_{w}} \right)}},
			\end{equation}
			\begin{equation}
				\label{Eq:TMSV_Radar_Parameter_2}
				\mathcal{A}_2^{\text{TMSV}} = 2 \sqrt{\frac{\left( \gamma n_{w} + \eta^s \right) K}{8 \gamma n_{w} + 7 \eta^s + 2 n_{w} + 1 + \frac{\eta^s}{\gamma}\left( 1 + \frac{1}{n_{w}} \right)}},
			\end{equation}
			where $\gamma = \eta^s n_s / n_w$ is defined identically to the CS radar model.
			
			\begin{figure}[!t]
				\centering
				\includegraphics[width=0.48\textwidth]{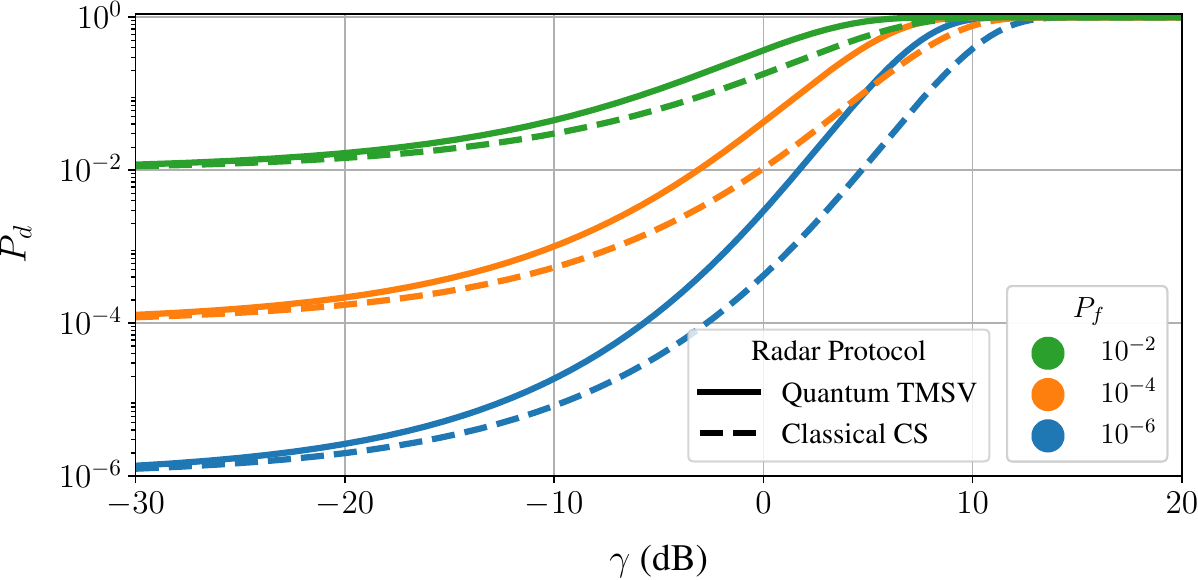}
				\caption{Detection probability versus SINR for the quantum TMSV and classical CS radar models.}
				\label{Fig:Pd_Vs_SINR_Classical_Quantum_Protocol}
			\end{figure}
			
			In quantum information theory, the fundamental performance limit for binary quantum hypothesis testing is given by the Helstrom bound~\cite{helstrom1969quantum}, which characterizes the minimum achievable error probability under optimal joint measurements. While this bound provides insight into the ultimate quantum gain limit of $6$~dB, the radar model considered in this work is based on the quantum TMSV protocol, which employs a suboptimal joint measurement scheme using a correlation detector. Although the maximum achievable advantage of the quantum TMSV radar is $3$~dB, it remains more practically feasible to implement than a conventional QI radar~\cite{bischeltsrieder2025note}. Moreover, radar performance is conventionally assessed using ROC curves, which specify detection and FA probabilities without requiring prior probabilities on the hypotheses. This ROC-based formulation enables a direct mapping of $P_d^{\text{min}}$ and $P_f^{\text{max}}$ requirements to an equivalent SINR threshold for ISAC optimization. For this reason, while acknowledging the Helstrom bound as the theoretical limit, we adopt the Gaussian detection model and the resulting ROC expressions for the radar protocols considered in this work.
			
			\begin{figure}[!t]
				\centering
				\includegraphics[width=0.48\textwidth]{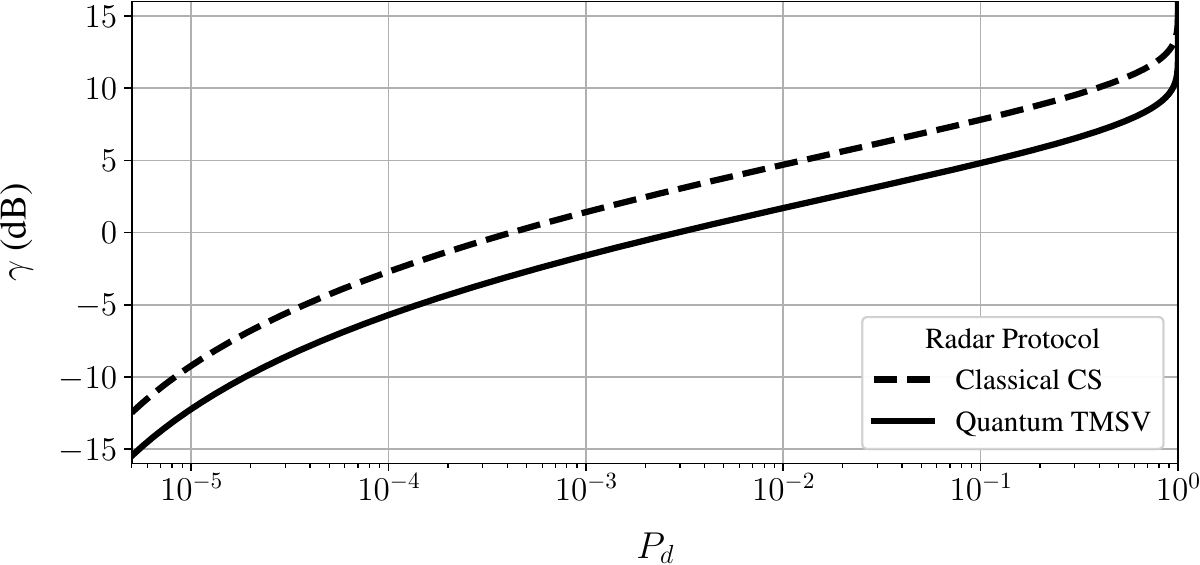}
				\caption{Required SINR versus $P_d$ for the quantum TMSV and CS radar models at $P_f = 10^{-6}$.}
				\label{Fig:SINR_Vs_Pd_Classical_Quantum_Protocol}
			\end{figure}
			
			Assuming a maximized quantum advantage of $\mathcal{Q} = 3$~dB, Fig.~\ref{Fig:Pd_Vs_SINR_Classical_Quantum_Protocol} shows the detection probability as a function of SINR for the quantum TMSV radar compared with its classical CS benchmark. It is evident that the quantum TMSV radar outperforms the CS radar over a wide SINR range, with the performance gap becoming more pronounced as $P_f$ decreases. In practical radar systems, a target $P_d$ is typically specified for a given $P_f$, which in turn determines the required SINR. To reflect this operational perspective, Fig.~\ref{Fig:SINR_Vs_Pd_Classical_Quantum_Protocol} shows the required SINR as a function of $P_d$ for $P_f = 10^{-6}$, clearly illustrating the quantum advantage, namely a reduction in the required SINR across varying detection probability requirements.
			
\begin{figure}[!t]
				\centering
				\includegraphics[width=0.48\textwidth]{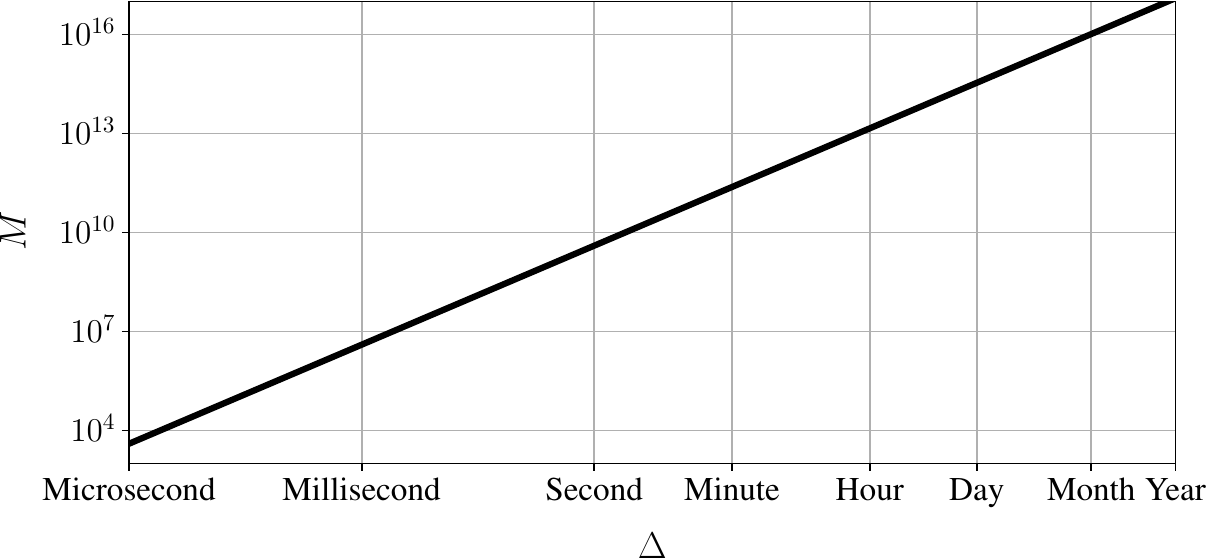}
				\caption{Number of independent modes as a function of dwell time for a TMSV source with bandwidth $B = 4.0$~GHz.}
				\label{Fig:Modes_Vs_Dwell_Time}
			\end{figure}

			The performance advantage of the quantum TMSV radar arises when the mean photon number per temporal--spectral mode, $n_{s,m}$, is much smaller than unity and the background thermal noise is high. For $f = 16.0$~GHz at room temperature ($T = 293.0$~K), the latter condition is satisfied with $n_w \approx 381.1$, as obtained from~(\ref{Eq:Bose-Einstein_Distribution}). To satisfy the low signal power requirement of quantum TMSV, the total signal energy, $\mathscr{E}_s$, must be distributed over a large number of temporal--spectral modes
			\begin{equation}
				\label{Eq:Energy_per_Mode}
				\mathscr{E}_{s,m} = \frac{\mathscr{E}_s}{M} = h f n_{s,m},
			\end{equation}
			where $\mathscr{E}_{s,m}$ is the energy per mode, $M$ is the number of modes, and $h$ is Planck's constant. The total number of independent modes is given by~\cite{shapiro2020quantum}
			\begin{equation}
				\label{Eq:Modes_Count}
				M = \Delta  B,
			\end{equation}
			where $\Delta$ is the signal dwell time and $B$ is the signal bandwidth. In the microwave regime ($3 < f < 30$~GHz), state-of-the-art Josephson parametric amplifiers can generate broadband TMSV states with bandwidths of $B = 3.0$~GHz~\cite{livreri2023microwave, macklin2015near}, and recent work has reported bandwidths up to $B = 4.0$~GHz~\cite{fasolo2022bimodal}. This leads to a mode count that grows linearly with dwell time, as shown in Fig.~\ref{Fig:Modes_Vs_Dwell_Time}.
			
			For a given SINR budget, $\gamma_b$, the SINR can be distributed across $M$ modes. From~(\ref{Eq:CS_Radar_Parameters}), when $K = M$ and $\gamma = \gamma_b / M$, the dependence on $M$ cancels out, so the classical CS radar performance is unaffected by $M$. In contrast, the quantum TMSV radar performance depends explicitly on $M$, as indicated by~(\ref{Eq:TMSV_Radar_Parameter_1}) and~(\ref{Eq:TMSV_Radar_Parameter_2}). The auxiliary parameters then take the form
			\begin{equation}
				\mathcal{A}_1^{\text{CS}} = 1, \qquad \mathcal{A}_2^{\text{CS}} = 2 \sqrt{\frac{\gamma_b n_w}{2 n_w + 1}},
			\end{equation}
			\begin{equation}
				\mathcal{A}_1^{\text{TMSV}} = \sqrt{1 + \frac{ \frac{4 \gamma_b n_w}{M} + 3 \eta^s }{2 n_w + 1+ \frac{\eta^s M}{\gamma_b} \left( 1 + \frac{1}{n_w} \right)}},
			\end{equation}
			\begin{equation}
				\hspace{-1.5mm} \mathcal{A}_2^{\text{TMSV}} \hspace{-1mm} = 2 \sqrt{\hspace{-0.5mm} \frac{\gamma_b n_w + \eta^s M}{\frac{8 \gamma_b n_w}{M} + 7 \eta^s + 2 n_w + 1 + \frac{\eta^s M}{\gamma_b}\left( 1 + \frac{1}{n_w} \right)}}. \hspace{-1mm}
			\end{equation}
			For the quantum TMSV radar, $M$ must be chosen such that the low photon number condition on $n_{s,m}$ is satisfied~\cite{sorelli2021detecting, pavan2024range}. Using~(\ref{Eq:Energy_per_Mode}) and~(\ref{Eq:Modes_Count}), the total signal power for $n_{s,m} = 0.5$ is evaluated as
			\begin{equation}
				\mathscr{P}_s = \frac{\mathscr{E}_s}{\Delta} = B h f n_{s,m} \approx 21.2 \text{ fW},
			\end{equation}
			with the corresponding spectral power density
			\begin{equation}
				\frac{\mathscr{P}_s}{B} \approx 5.3 \times 10^{-24} \text{ W/Hz},
			\end{equation}
			which equals the energy per mode, $\mathscr{E}_{s,m}$.
			
\begin{figure}[!t]
				\centering
				\includegraphics[width=0.48\textwidth]{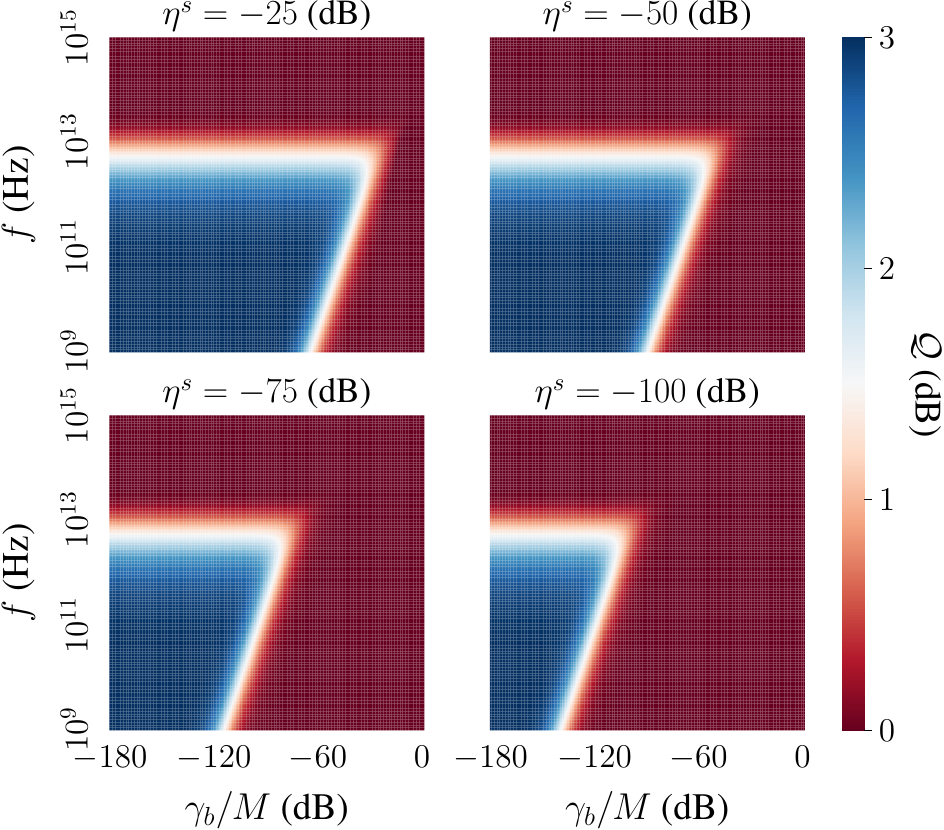}
				\caption{Quantum advantage as a function of channel transmissivity and SINR per mode.}
				\label{Fig:Quantum_Adv_Heatmap}
			\end{figure}

\begin{figure}[!t]
				\centering
				\includegraphics[width=0.48\textwidth]{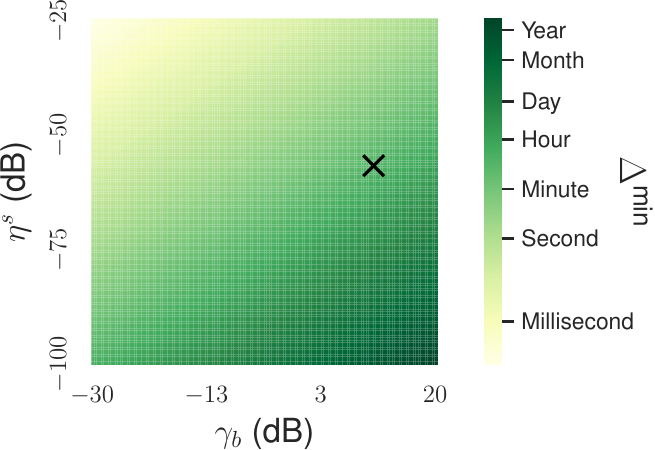}
				\caption{Minimum required dwell time as a function of channel transmissivity and SINR, with the cross indicating the operating point used in this work.}
				\label{Fig:Dwell_Time_Heatmap}
			\end{figure}

			The achievable quantum advantage, $0 \le \mathcal{Q} \le 3$~dB, depends on the signal frequency, $f$, SINR per mode, $\gamma_b / M$, and channel transmissivity, $\eta^s$. Fig.~\ref{Fig:Quantum_Adv_Heatmap} shows how $\mathcal{Q}$ varies with these parameters. At high signal frequencies, the quantum advantage vanishes because the number of background noise photons approaches zero, as shown in Fig.~\ref{Fig:NoisePhotons_Vs_Wavelength}. For a fixed signal frequency, increasing channel loss reduces the maximum allowable SINR per mode needed to attain full quantum advantage. This necessitates larger $M$, and thus longer dwell times. For $B = 4.0$~GHz and $f = 16.0$~GHz, Fig.~\ref{Fig:Dwell_Time_Heatmap} shows the minimum required dwell time as a function of $\eta^s$ and $\gamma_b$. The dark region corresponds to dwell times that are practically unfavorable for the considered quantum TMSV radar configuration. In fast-varying target or channel conditions, the dwell time, $\Delta$, must lie within the scene's stationarity interval. When this interval is short, $\Delta$ can be reduced by relaxing the sensing requirements (leading to lower SINR), operating with higher channel transmissivity, $\eta^s$, or using a larger system bandwidth, $B$, all of which reduce the number of modes, $M$, required to reach the quantum advantage regime.
			
			The quantum TMSV radar analysis in this work assumes ideal quantum hardware, without accounting for practical non-idealities such as limited entanglement fidelity, quantum memory storage loss, detector inefficiency, and other implementation imperfections that may reduce the achievable quantum advantage. A detailed overview of these system-level considerations can be found in~\cite{pan2024evolution}, which offers useful guidance for developing more realistic quantum TMSV radar models.
\section{Integrated Quantum Sensing and Classical Communication (IQSCC) System}~\label{Sec:IQSCC_System}
	In the proposed IQSCC system, a quantum TMSV radar is employed for target sensing. The radar SINR constraint in~(\ref{Eq:Classical_Radar_SINR_Constraint}) is not directly applicable to the quantum TMSV case, because the total transmit covariance, $\mathbf{V}_{\hspace{-0.5mm} t}$, contains power allocated to both the sensing and communication signals. In the IQSCC framework, only the sensing covariance, $\mathbf{V}_{\hspace{-0.5mm} s}$, contributes to the quantum TMSV detection process, while the communication covariance, $\mathbf{V}_{\hspace{-0.5mm} c}$, produces additional interference after being reflected from the target. As a result, the effective radar SINR is given by
	\begin{equation}
		\mathbf{a}_r^H(\theta_0)\mathbf{\Psi}_q^{-1} \mathbf{a}_r(\theta_0) \ge \frac{\rho^s}{|\beta_0|^2} \left[\mathbf{a}_t^H(\theta_0) \mathbf{V}_{\hspace{-0.5mm} s} \mathbf{a}_t(\theta_0) \right]^{-1},
	\end{equation}
	where $\mathbf{\Psi}_q = \mathbf{\Psi} + |\beta_0|^2 \mathbf{A}(\theta_0) \mathbf{V}_{\hspace{-0.5mm} c} \mathbf{A}^H(\theta_0)$. In the conventional ISAC baseline, both the sensing and communication signals contribute to target detection, so this additional interference term does not arise. The utilized parameters to compare the performance of the conventional ISAC and proposed IQSCC systems are shown in Table~\ref{Tab:Parameters}. The BS is equipped with $N_r = N_t = 10$ receive and transmit antennas, respectively. The maximum power budgets are set to $P_b^{\text{max}} = 1.0$~W and $P_u^{\text{max}} = 200.0$~mW for the BS and UL user, respectively. The target reflection coefficient is $|\beta_0|^2 = \eta^s = -56.5$~dB. The path loss is $\eta^d = \eta^u = -60.0$~dB for both DL and UL users, $\eta^i = -20.0$~dB for the interferers, and the residual SI channel power is $P_r = -100.0$~dB. The background noise power is computed as follows~\cite{pavan2024range}
	\begin{equation}
		\sigma_w^2 = n_{w} h f B \approx 16.2 \text{ pW}.
	\end{equation}
	Considering the requirement on the FA and detection probabilities to be $P_f^{\text{max}} = 10^{-6}$ and $P_d^{\text{min}} = 0.99$, respectively, the radars' minimum SINR requirement is obtained from Fig.~\ref{Fig:SINR_Vs_Pd_Classical_Quantum_Protocol} as $\rho^{\text{CS}} = 14.0$~dB and $\rho^{\text{TMSV}} = 11.0$~dB, where the SINR per mode requirement on the IQSCC system is $\rho^{\text{TMSV}} / M \approx -112.0$~dB.
	
	\def\leftvspace{\rule{0pt}{4mm}}
\def\rightvspace{0.7mm}

\begin{table}[!t]
	\centering
	\caption{The utilized ISAC and quantum TMSV radar system parameters.}
	\label{Tab:Parameters}
	\resizebox{0.495\textwidth}{!}{
		\begin{tabular}{|c|c|c|}
			\hline
			\leftvspace \textbf{Parameter} & \textbf{Value} & \textbf{Dependency} \\ [\rightvspace] \hline
			
			\leftvspace Photons per Mode & $n_{s,m} = 0.5$ & \begin{tabular}[c]{@{}c@{}} \leftvspace Theory \\[0.5mm] Limitations \\[\rightvspace] \end{tabular} \\ [\rightvspace] \hline
			
			\leftvspace Operating Temperature & $T = 293.0$ K & \begin{tabular}[c]{@{}c@{}} \leftvspace Environmental \\[0.5mm] Limitations \\[\rightvspace] \end{tabular} \\ [\rightvspace] \hline
			
			\leftvspace Signal Frequency & $f = 16.0$ GHz & \multirow{2}{*}{\vspace{-1mm}\begin{tabular}[c]{@{}c@{}} Practical \\[0.5mm] Limitations \end{tabular}} \\ [\rightvspace] \cline{1-2}
			
			\leftvspace Signal Bandwidth & $B = 4.0$ GHz & \\ [\rightvspace] \hline
			
			\leftvspace Thermal Noise Photons & $n_w \approx 381.1$ & $T$ and $f$ \\ [\rightvspace] \hline
			
			\leftvspace AWGN Power & $\sigma_w^2 \approx 16.2$ pW & $f$ and $B$ \\ [\rightvspace] \hline
			
			\leftvspace FA Probability & $P_f^{\text{max}} = 10^{-6}$ & \multirow{2}{*}{\vspace{-1mm}\begin{tabular}[c]{@{}c@{}} Performance \\[0.5mm] Requirement \end{tabular}} \\ [\rightvspace] \cline{1-2}
			
			\leftvspace Detection Probability & $P_d^{\text{min}} = 0.99$ & \\ [\rightvspace] \hline
			
			\leftvspace CS Radar SINR & $\rho^{\text{CS}} = 14.0$ dB & \multirow{2}{*}{\vspace{-1mm}$P_f^{\text{max}}$ and $P_d^{\text{min}}$} \\ [\rightvspace] \cline{1-2}
			
			\leftvspace TMSV Radar SINR & $\rho^{\text{TMSV}} = 11.0$ dB & \\ [\rightvspace] \hline
			
			\leftvspace BS--Target Channel Gain & $\eta^s = -56.5$ dB & \multirow{2}{*}{\vspace{-10.5mm}\begin{tabular}[c]{@{}c@{}}Environmental \\[0.5mm] Limitations\end{tabular}} \\ [\rightvspace] \cline{1-2}
			
			\leftvspace BS--DL User Channel Gain  & $\eta^d = -60.0$ dB & \\ [\rightvspace] \cline{1-2}
			
			\leftvspace BS--UL User Channel Gain & $\eta^u = -60.0$ dB & \\ [\rightvspace] \cline{1-2}
			
			\leftvspace BS--Interferer Channel Gain & $\eta^i = -20.0$ dB & \\ [\rightvspace] \hline
			
			\leftvspace Dwell Time & $\Delta \approx 502.4$ s & $\gamma$, $\eta$, and $B$ \\ [\rightvspace] \hline
			
			\leftvspace Modes Count & $M \approx 2.0 \times 10^{12}$ & $\Delta$ and $B$ \\ [\rightvspace] \hline
			
			\leftvspace Receive Antennas Count & $N_r = 10$ & \multirow{4}{*}{\vspace{-11mm}\begin{tabular}[c]{@{}c@{}} Practical \\[0.5mm] Limitations \end{tabular}} \\ [\rightvspace] \cline{1-2}
			
			\leftvspace Transmit Antennas Count & $N_t = 10$ & \\ [\rightvspace] \cline{1-2}
			
			\leftvspace BS Power Budget & $P_b^{\text{max}} = 1.0$ W & \\ [\rightvspace] \cline{1-2}
			
			\leftvspace UL User Power Budget & $P_u^{\text{max}} = 200.0$ mW & \\ [\rightvspace] \cline{1-2}
			
			\leftvspace Residual SI Channel Power & $P_r = -100.0$ dB & \\ [\rightvspace] \hline
		\end{tabular}
	}
\end{table}

\begin{figure}[!t]
		\centering
		\subfloat[]{
			\includegraphics[width=0.48\textwidth]{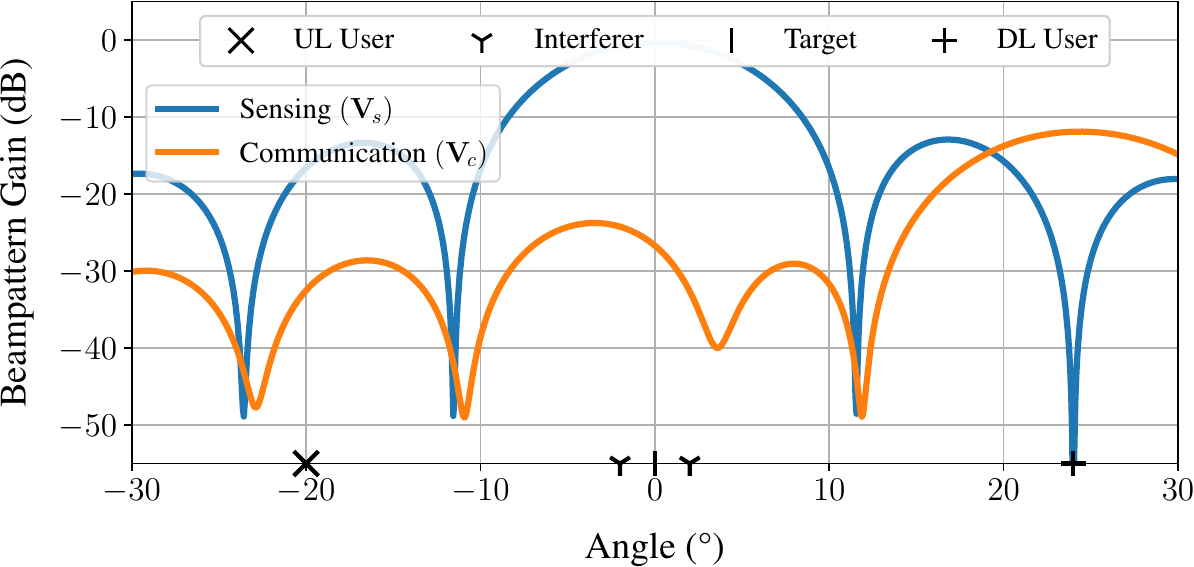}
		}
		\vfil
		\subfloat[]{
			\includegraphics[width=0.48\textwidth]{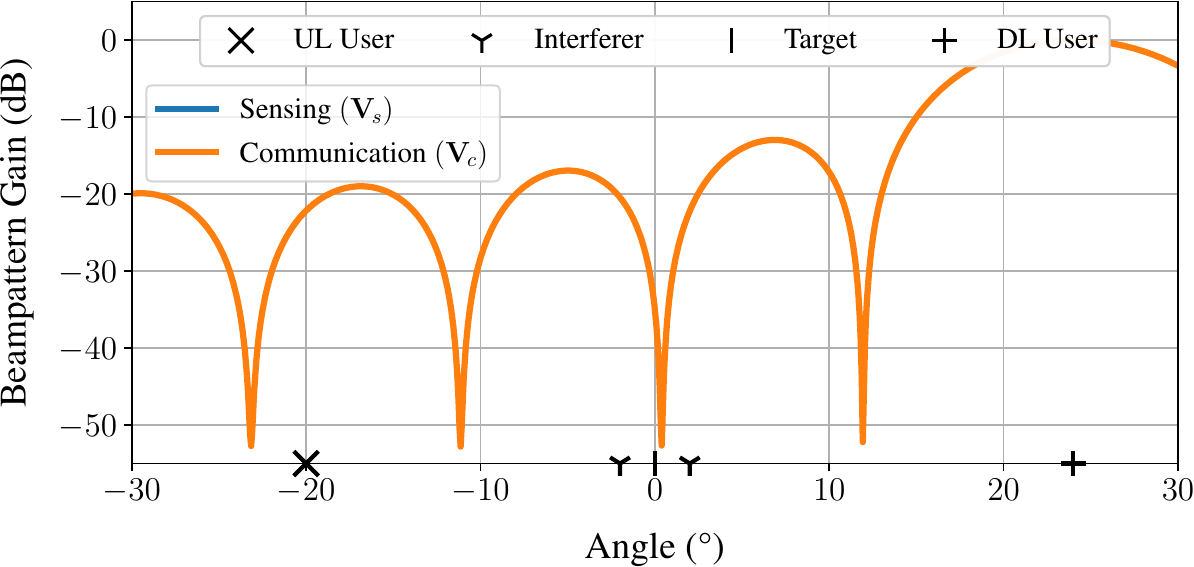}
		}
		\caption{Beampattern gain for the (a) conventional ISAC $(\rho^{\text{CS}} = 14.0$~dB$)$ and (b) proposed IQSCC $(\rho^{\text{TMSV}} / M \approx -112.0$~dB$)$ systems.}
		\label{Fig:Beampattern_Gain}
	\end{figure}

	Both modeled systems employ identical parameters and the same SCA-based joint beamforming and power-allocation algorithm. Thus, any performance improvement of the IQSCC system stems solely from the quantum-enhanced sensing advantage, which relaxes the radar SINR constraint and therefore allows more transmit power to be allocated to the communication signal. The beampattern gains for both systems are illustrated in Fig.~\ref{Fig:Beampattern_Gain}. For the conventional ISAC system, the sensing beam exhibits a sharp mainlobe in the target direction while remaining highly attenuated toward the DL user, demonstrating strong spatial isolation and minimal interference with the communication functionality. In contrast, the communication beam forms a pronounced peak toward the DL user, with sidelobes suppressed in the target direction. Although both systems achieve the same sensing performance, the IQSCC configuration in Fig.~\ref{Fig:Beampattern_Gain}~(b) does so with reduced overall sensing power, with the absence of a visible sensing beampattern due to the extremely low SINR per mode.
	
	\begin{figure}[!t]
		\centering
		\includegraphics[width=0.48\textwidth]{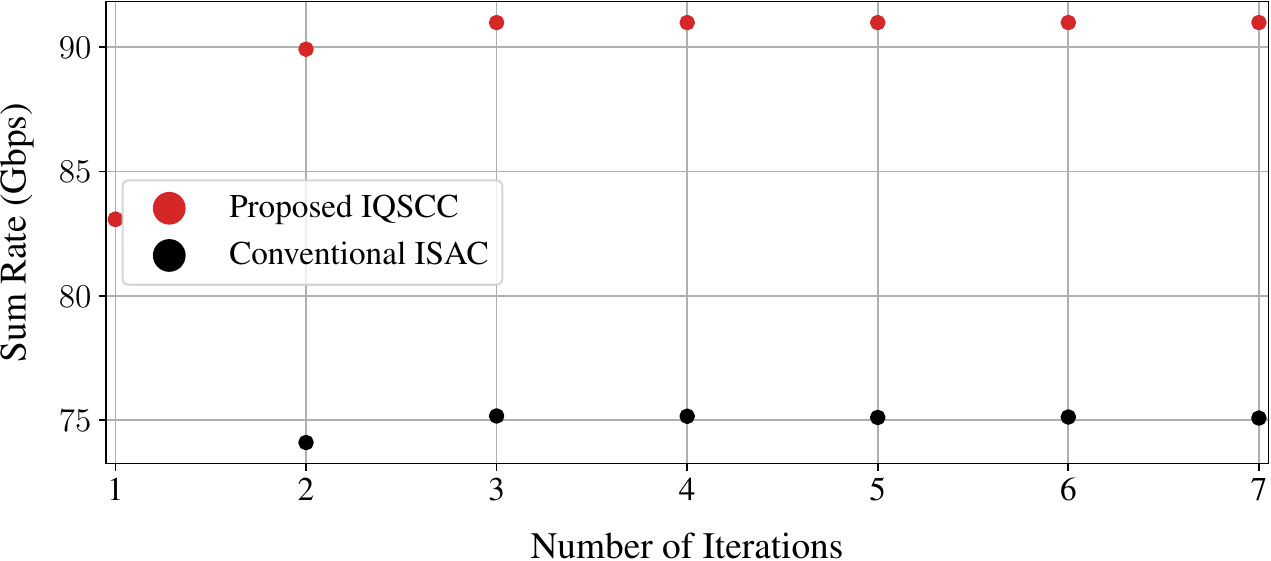}
		\caption{Achieved sum rate versus iteration index for the conventional ISAC and proposed IQSCC systems.}
		\label{Fig:Sum_Rate_Vs_Iterations}
	\end{figure}
	
	The system sum rates over successive optimization iterations are presented in Fig.~\ref{Fig:Sum_Rate_Vs_Iterations}. As shown, the proposed IQSCC system converges rapidly to a higher steady-state sum rate than the conventional ISAC baseline, achieving approximately $91.0$~gigabits per second (Gbps). This represents a $21.2\%$ improvement, corresponding to an increase of $15.9$~Gbps under the same bandwidth and power conditions. This gain highlights the advantage of incorporating quantum sensing into the ISAC framework, which allows for more efficient utilization of power and spatial resources.
\section{Conclusion}~\label{Sec:Conclusion}
	This paper presented an IQSCC architecture that embeds quantum TMSV radar within an FD ISAC system. A detection framework was developed for both the quantum TMSV and classical radar benchmarks, which was used to demonstrate the superior detection performance of quantum TMSV in low-SINR regimes. A joint optimization problem incorporating beamforming and power allocation was formulated and solved using SCA techniques. Simulation results showed that the IQSCC system increased the overall sum rate by $15.9$~Gbps compared to conventional ISAC systems while satisfying the required sensing constraints.
	
	Future research may explore experimental validation, extensions to multi-user scenarios, and hybrid quantum--classical detection strategies in which the reflected classical communication signal is exploited rather than treated solely as interference. Such approaches may further enhance sensing performance and broaden the applicability of the proposed framework. Another promising direction is the development of quantum TMSV radar models that explicitly capture temporal effects across multiple observations, including phase drift, idler-storage decay, and time-varying hardware impairments. Incorporating these dynamics would enable analysis of coherent versus non-coherent integration over time and allow the sensing model to more accurately reflect physical hardware behavior and better inform ISAC system optimization.
\appendix
\subsection{Derivation of~(\ref{Eq:Receive_Beamformer_u}) and~(\ref{Eq:Receive_Beamformer_w})}~\label{App:Receive_Beamformers}
	The optimal receive beamformers can be derived in closed form using the generalized Rayleigh quotient
	\begin{equation}
		R(\mathbf{x}) = \frac{\mathbf{x}^H \mathbf{M}_n \mathbf{x}}{\mathbf{x}^H \mathbf{M}_d \mathbf{x}},
	\end{equation}
	where $\mathbf{x}$ is a non-zero complex vector, $\mathbf{M}_n$ is a complex Hermitian matrix, and $\mathbf{M}_d$ is a positive-definite complex Hermitian matrix such that $\mathbf{x}^H \mathbf{M}_d \mathbf{x} > 0$ for all $\mathbf{x} \neq 0$. The maximum value of $R(\mathbf{x})$ occurs at the largest generalized eigenvalue, $\lambda$, of the pair $(\mathbf{M}_n, \mathbf{M}_d)$
	\begin{equation}
		\mathbf{M}_n \mathbf{x} = \lambda \mathbf{M}_d \mathbf{x}.
	\end{equation}
	To optimize the receive beamformer $\mathbf{u}$,~(\ref{Eq:Radar_SINR_1}) is rewritten as~\cite{he2023full}
	\begin{equation}
		\begin{split}
			\gamma^s &= 
			\frac{|\beta_0|^2 \mathbf{u}^H \mathbf{a}_r(\theta_0) \mathbf{a}_t^H(\theta_0) \mathbf{V}_{\hspace{-0.5mm} t} \mathbf{a}_t(\theta_0) \mathbf{a}_r^H(\theta_0) \mathbf{u}}
			{\mathbf{u}^H ( p \mathbf{h}\mathbf{h}^H + \mathbf{B} \mathbf{V}_{\hspace{-0.5mm} t} \mathbf{B}^H + \sigma_w^2 \mathbf{I}_{N_r} ) \mathbf{u}} \\[2mm]
			&= \frac{|\beta_0|^2 \mathbf{a}_t^H(\theta_0) \mathbf{V}_{\hspace{-0.5mm} t} \mathbf{a}_t(\theta_0) \mathbf{u}^H \mathbf{a}_r(\theta_0) \mathbf{a}_r^H(\theta_0)   \mathbf{u}}{\mathbf{u}^H ( p \mathbf{h}\mathbf{h}^H + \mathbf{B} \mathbf{V}_{\hspace{-0.5mm} t} \mathbf{B}^H + \sigma_w^2 \mathbf{I}_{N_r} ) \mathbf{u}},
		\end{split}
	\end{equation}
	since $|\beta_0|^2 \mathbf{a}_t^H(\theta_0) \mathbf{V}_{\hspace{-0.5mm} t} \mathbf{a}_t(\theta_0)$ is a scalar. Comparing with the generalized Rayleigh quotient, by letting $\mathbf{x} = \mathbf{u}$, it follows that $\mathbf{M}_n = |\beta_0|^2 \mathbf{a}_t^H(\theta_0) \mathbf{V}_{\hspace{-0.5mm} t} \mathbf{a}_t(\theta_0) \mathbf{a}_r(\theta_0) \mathbf{a}_r^H(\theta_0)$ and $\mathbf{M}_d = p\mathbf{h}\mathbf{h}^H + \mathbf{B} \mathbf{V}_{\hspace{-0.5mm} t} \mathbf{B}^H + \sigma_w^2 \mathbf{I}_{N_r}$. Therefore, the generalized eigenvalue problem is given by
	\begin{equation}
		\begin{split}
			|\beta_0|^2 \mathbf{a}_t^H(\theta_0) \mathbf{V}_{\hspace{-0.5mm} t} &\mathbf{a}_t(\theta_0)\mathbf{a}_r(\theta_0) \mathbf{a}_r^H(\theta_0)\mathbf{u} \\
			&= \lambda (p\mathbf{h}\mathbf{h}^H + \mathbf{B} \mathbf{V}_{\hspace{-0.5mm} t} \mathbf{B}^H + \sigma_w^2 \mathbf{I}_{N_r}) \mathbf{u}.
		\end{split}
	\end{equation}
	Since $\mathbf{a}_r^H(\theta_0)\mathbf{u}$ is scalar, this expression can be rewritten as
	\begin{equation}
		(p\mathbf{h}\mathbf{h}^H + \mathbf{B} \mathbf{V}_{\hspace{-0.5mm} t} \mathbf{B}^H + \sigma_w^2 \mathbf{I}_{N_r})^{-1} \mathbf{a}_r(\theta_0) = \frac{\lambda}{\nu} \mathbf{u},
	\end{equation}
	where $\nu = |\beta_0|^2 \mathbf{a}_t^H(\theta_0) \mathbf{V}_{\hspace{-0.5mm} t} \mathbf{a}_t(\theta_0) \mathbf{a}_r^H(\theta_0)\mathbf{u}$. Maximizing the SINR corresponds to the largest eigenvalue, and thus the optimal receive beamformer for the radar is obtained as~(\ref{Eq:Receive_Beamformer_u}). The optimal receive beamformer for~(\ref{Eq:Uplink_SINR}) is derived similarly to obtain~(\ref{Eq:Receive_Beamformer_w}).
\subsection{Derivation of~(\ref{Eq:ROC})}~\label{App:ROC}
	For a fixed $P_f$, $P_d$ is maximized using the likelihood ratio test~\cite{kay1993fundamentals}
	\begin{equation}
		\label{Eq:LRT_1}
		L(\mathbf{r}) = \frac{p( \mathbf{r} | \mathcal{H}_1)}{p( \mathbf{r} | \mathcal{H}_0)} > \zeta,
	\end{equation}
	where $\mathbf{r}$ is the received signal vector and $\zeta$ is a decision threshold determined to satisfy
	\begin{equation}
		\int_{\left\{\mathbf{r} : L(\mathbf{r}) > \zeta \right\}} p( \mathbf{r} | \mathcal{H}_0) \; d\mathbf{r} = P_f.
	\end{equation}
	Assuming Gaussian statistics, the likelihoods under each hypothesis are
	\begin{equation}
		\begin{split}
			p( \mathbf{r} | \mathcal{H}_0) &= \prod_{k = 1}^K \frac{e^{-\frac{\left(\mathbf{r}^{[k]}\right)^2}{2\sigma_0^2}}}{\sigma_0 \sqrt{2\pi}} = \frac{e^{-\frac{\sum_{k = 1}^K \left(\mathbf{r}^{[k]}\right)^2}{2\sigma_0^2}}}{\left(\sigma_0 \sqrt{2\pi}\, \right)^K},
		\end{split}
	\end{equation}
	\begin{equation}
		\begin{split}
			p( \mathbf{r} | \mathcal{H}_1) &= \prod_{k = 1}^K \frac{e^{-\frac{\left( \mathbf{r}^{[k]} - \mu_1 \right)^2}{2\sigma_1^2}}}{\sigma_1 \sqrt{2\pi}} = \frac{e^{-\frac{\sum_{k = 1}^K \left( \mathbf{r}^{[k]} - \mu_1 \right)^2}{2\sigma_1^2}}}{\left( \sigma_1 \sqrt{2\pi} \, \right)^K}.
		\end{split}
	\end{equation}
	Substituting these likelihoods into~(\ref{Eq:LRT_1}) yields the decision rule
	\begin{equation}
		\label{Eq:LRT_2}
		\begin{split}  
			& \left( \frac{\sigma_0}{\sigma_1} \right)^K e^{-\frac{\sum_{k = 1}^K \left( \mathbf{r}^{[k]} - \mu_1 \right)^2}{2\sigma_1^2} + \frac{\sum_{k = 1}^K \left( \mathbf{r}^{[k]} \right)^2}{2\sigma_0^2}} > \zeta \\[2mm]
			& \hspace{-3mm} \Rightarrow \sum_{k = 1}^K \left[ \frac{ \left( \mathbf{r}^{[k]} \right)^2}{\sigma_0^2} - \frac{\left( \mathbf{r}^{[k]} - \mu_1 \right)^2}{\sigma_1^2} \right] > 2 \ln\left(\zeta \left( \frac{\sigma_1}{\sigma_0} \right)^K \right).
		\end{split}
	\end{equation}
	Expanding and simplifying the left-hand side of~(\ref{Eq:LRT_2}) leads to
	\begin{equation}
		\begin{split}
			\left(\frac{1}{\sigma_0^2} - \frac{1}{\sigma_1^2}\right)\sum_{k = 1}^K \left( \mathbf{r}^{[k]} \right)^2 + & \frac{2\mu_1}{\sigma_1^2} \sum_{k = 1}^K \mathbf{r}^{[k]} > \\
			& \hspace{1mm} 2 \ln\left(\zeta \left( \frac{\sigma_1}{\sigma_0} \right)^K \right) + \frac{K \mu_1^2}{\sigma_1^2}.
		\end{split}
	\end{equation}
	Defining the sample mean as
	\begin{equation}
		\mathbf{\bar{r}} = \frac{1}{K} \sum_{k = 1}^K \mathbf{r}^{[k]},
	\end{equation}
	the test can be rewritten as
	\begin{equation}
		\hspace{-2mm} \mathbf{\bar{r}} > \underbrace{\hspace{-0.5mm} \frac{\sigma_1^2 \ln\left(\hspace{-0.5mm} \zeta \left( \frac{\sigma_1}{\sigma_0} \right)^K \right)}{K \mu_1} + \hspace{-0.5mm} \frac{\mu_1}{2} \hspace{-0.5mm} + \hspace{-0.5mm} \frac{\left(1 - \frac{\sigma_1^2}{\sigma_0^2}\right)\sum_{k = 1}^K \left( \mathbf{r}^{[k]} \right)^2}{2 K \mu_1}\hspace{-0.5mm}}_{\zeta'},
	\end{equation}
	which, together with the test statistic in~(\ref{Eq:Test_Statistic}), defines the FA and detection probabilities leading to ~(\ref{Eq:ROC}).
	\subsection{Derivation of~(\ref{Eq:CS_Radar_Parameters})}~\label{App:CS_Radar_Parameters}
		The CS radar model is based on quantum optical formalism, where quantum states of light are described using annihilation, $\hat{a}$, and creation, $\hat{a}^\dagger$, operators. Their action on the Fock basis states, $\ket{n}$, is given by~\cite{weedbrook2012gaussian}
		\begin{equation}
			\hat{a} \ket{n} = \sqrt{n} \ket{n - 1},
		\end{equation}
		\begin{equation}
			\hat{a}^{\dagger} \ket{n} = \sqrt{n + 1} \ket{n + 1}.
		\end{equation}
		In the shot-noise unit convention, the annihilation and creation operators can be expressed in terms of the dimensionless quadrature operators, $\hat{x}$ and $\hat{p}$, as follows~\cite{weedbrook2012gaussian}
		\begin{equation}
			\label{Eq:Annihilation_Operator}
			\hat{a} = \frac{1}{2} \left(\hat{x} + j \hat{p} \right),
		\end{equation}
		\begin{equation}
			\label{Eq:Creation_Operator}
			\hat{a}^{\dagger} = \frac{1}{2} \left(\hat{x} - j \hat{p} \right).
		\end{equation}
		Accordingly, the quadrature operators are recoverable from the bosonic operators via the following relation
		\begin{equation}
			\label{Eq:x_quadrature}
			\hat{x} = \hat{a} + \hat{a}^{\dagger},
		\end{equation}
		\begin{equation}
			\hat{p} = \frac{1}{j} \left( \hat{a} - \hat{a}^{\dagger} \right).
		\end{equation}
		The photon number operator is defined as~\cite{weedbrook2012gaussian}
		\begin{equation}
			\hat{n} = \hat{a}^{\dagger} \hat{a},
		\end{equation}
		with its expectation in a CS $\ket{\alpha}$ giving the average photon number
		\begin{equation}
			n = \left \langle \alpha \middle| \hat{n} \middle| \alpha \right \rangle.
		\end{equation}
		The average signal power per pulse is then
		\begin{equation}
			\label{Eq:Signal_Power}
			\langle P \rangle = \frac{h f}{\delta} n,
		\end{equation}
		where $h$ is Planck's constant, $f$ is the signal frequency, and $\delta$ is the pulse duration. In practice, the usable signal bandwidth, $\tilde{W}$, is constrained by the phase-matching bandwidth, $W$, such that $\tilde{W} \le W$. To prevent spectral spreading beyond this limit, the time duration $\delta$ must satisfy $\delta \gtrsim W$~\cite{dorrer2021optical}. Substituting this into~(\ref{Eq:Signal_Power}) yields
		\begin{equation}
			\left \langle P \right \rangle = h f W n.
		\end{equation}
		From~(\ref{Eq:Quantum_Channel_Received_Signal}), the expected photon number at the receiver is
		\begin{equation}
			\label{Eq:Received_Photons_Count}
			\begin{split}
			n_r^{[k]}  &= \left \langle \left(\hat{a}_r^{[k]} \right)^{ \dagger} \hat{a}_r^{[k]} \right \rangle \\
			&= \eta^s n_t^{[k]}  + \left(1 - \eta^s \right) \left \langle \left( \hat{a}_n^{[k]} \right)^{\dagger} \hat{a}_n^{[k]} \right \rangle,
			\end{split}
		\end{equation}
		where $n_t^{[k]} = \left \langle \left( \hat{a}_t^{[k]} \right)^{\dagger} \hat{a}_t^{[k]} \right \rangle$ denotes the mean transmitted photon number, while the cross terms vanish due to statistical independence between signal and noise. To ensure fair hypothesis testing, the noise level at the receiver must remain fixed under both hypotheses. This is enforced by defining the noise photon number as
		\begin{equation}
			\label{Eq:Noise_Photons_Count}
			\left \langle \left( \hat{a}_n^{[k]} \right)^{\dagger} \hat{a}_n^{[k]} \right \rangle =
			\begin{cases}
				n_{w}^{[k]}, & \text{under } \mathcal{H}_0, \\
				\frac{n_{w}^{[k]}}{1 - \eta^s}, & \text{under } \mathcal{H}_1.
			\end{cases}
		\end{equation}
		The thermal photon number, $n_{w}$, follows the Bose--Einstein distribution~\cite{einstein1925quantentheorie}
		\begin{equation}
			\label{Eq:Bose-Einstein_Distribution}
			n_{w} = \left[e^{hf / (T k_b)} - 1\right]^{-1},
		\end{equation}
		where $T$ is the environment temperature and $k_b$ is the Boltzmann's constant. Substituting~(\ref{Eq:Noise_Photons_Count}) into~(\ref{Eq:Received_Photons_Count}) yields the expected received photon number as
		\begin{equation}
			\label{Eq:Quantum_Channel_Received_Photons}
			n_r^{[k]} = \eta^s n_t^{[k]} + n_{w}^{[k]}.
		\end{equation}
		From~(\ref{Eq:x_quadrature}), the $\hat{x}$ quadrature expectation of the transmitted signal is
		\begin{equation}
			\left \langle \hat{x}_t \right \rangle = \left \langle \alpha \middle| \hat{x}_t \middle| \alpha \right \rangle = \alpha + \alpha^* = 2 \alpha,
		\end{equation}
		where $\alpha \in \mathbb{R}$ is assumed (i.e., the signal encoded only in $\hat{x}$). The second moment is
		\begin{equation}
			\label{Eq:x_Quadrature_Second_Moment}
			\left \langle \hat{x}_t^2 \right \rangle = \alpha^2 + \left(\alpha^*\right)^2 + 1 + 2 |\alpha|^2 = 4 \alpha^2 + 1,
		\end{equation}
		where the following commutator property has been used
		\begin{equation}
			\left[\hat{a}, \hat{a}^{\dagger} \right] = \hat{a} \hat{a}^{\dagger} - \hat{a}^{\dagger} \hat{a} = 1.
		\end{equation}
		Thus, the quadrature variance is given by
		\begin{equation}
			\sigma_{\hat{x}_t}^2 = \left \langle \hat{x}_t^2 \right \rangle - \left \langle \hat{x}_t \right \rangle^2 = 1.
		\end{equation}
		Similarly, the $\hat{p}$ quadrature parameters are found to be
		\begin{equation}
			\label{Eq:p_Quadrature_Parameters}
			\left \langle \hat{p}_t \right \rangle = 0, \quad \left \langle \hat{p}_t^2 \right \rangle = 1, \quad \sigma_{\hat{p}_t}^2 = 1.
		\end{equation}
		The transmitted photon number can be re-expressed using~(\ref{Eq:x_Quadrature_Second_Moment})~and~(\ref{Eq:p_Quadrature_Parameters}) as follows
		\begin{equation}
			n_t = \left \langle \left(\hat{a}_t \right)^{ \dagger} \hat{a}_t \right \rangle = \frac{1}{4} \left( \left \langle \hat{x}_t^2 \right \rangle + \left \langle \hat{p}_t^2 \right \rangle \right) = n_s + \frac{1}{2},
		\end{equation}
		where $n_s = \alpha^2$ is the signal photon count. Thus, the CS radar SINR is
		\begin{equation}
			\label{Eq:Quantum_Channel_SINR}
			\gamma^{\text{CS}} = \frac{\eta^s n_s}{n_{w}}.
		\end{equation}
		For reference, the vacuum quadrature variances are
		\begin{equation}
			\label{Eq:Vacuum_Variance}
			\sigma_{\hat{x}_v}^2 = \sigma_{\hat{p}_v}^2 = 1,
		\end{equation}
		corresponding to a vacuum photon number of
		\begin{equation}
			n_v = \frac{1}{2}.
		\end{equation}
		For a thermal state with $n_{w}$ photons, the quadrature variances are
		\begin{equation}
			\sigma_{\hat{x}_n}^2 = \sigma_ {\hat{p}_n}^2 = 2 n_{w} + 1.
		\end{equation}
		At the receiver, the expected $\hat{x}$ quadrature under $\mathcal{H}_1$ is
		\begin{equation}
			\label{Eq:CS_Received_x_Quadrature_Mean}
			\left \langle \hat{x}_r \right \rangle = \sqrt{\eta^s} \left(\left \langle \hat{a}_t \right \rangle + \left \langle \hat{a}_t^{\dagger} \right \rangle \right) = 2 \sqrt{\eta^s} \alpha,
		\end{equation}
		where~(\ref{Eq:Quantum_Channel_Received_Signal}) and~(\ref{Eq:x_quadrature}) are used, along with the fact that $\left \langle \hat{a}_n \right \rangle = \left \langle \hat{a}_n^{\dagger} \right \rangle =  0$. The corresponding second moment is
		\begin{equation}
			\left \langle \hat{x}_r^2 \right \rangle = \eta^s \left \langle \hat{x}_t^2 \right \rangle + (1 - \eta^s)\left \langle \hat{x}_n^2 \right \rangle = 2 \left( 2\eta^s \alpha^2 + n_{w} \right) + 1,
		\end{equation}
		where we made use of~(\ref{Eq:Quantum_Channel_Received_Signal}),~(\ref{Eq:x_Quadrature_Second_Moment}),~(\ref{Eq:Vacuum_Variance}), and $\left \langle \hat{x}_n \right \rangle =  0$. Hence, the variance is
		\begin{equation}
			\label{Eq:CS_Received_x_Quadrature_Variance}
			\sigma_{\hat{x}_r}^2 = 2n_{w} + 1.
		\end{equation}
		By symmetry, the same holds for $\hat{p}$. Finally,~(\ref{Eq:CS_Radar_Parameters}) is obtained by considering~(\ref{Eq:Quantum_Channel_SINR}) while plugging~(\ref{Eq:CS_Received_x_Quadrature_Mean})~and~(\ref{Eq:CS_Received_x_Quadrature_Variance}) into~(\ref{Eq:Radar_Parameters}).
	\subsection{Derivation of~(\ref{Eq:TMSV_Radar_Parameter_1}) and~(\ref{Eq:TMSV_Radar_Parameter_2})}~\label{App:TMSV_Radar_Parameters}
		The TMSV state has a zero-mean Gaussian Wigner representation with a covariance matrix in standard form given by~\cite{weedbrook2012gaussian}
		\begin{equation}
			\mathbf{V}_q = 
			\begin{bmatrix}
				S & 0 & C_q & 0 \\
				0 & S & 0 & -C_q \\
				C_q & 0 & S & 0 \\
				0 & -C_q & 0 & S
			\end{bmatrix},
		\end{equation}
		where $S = 2 n_q + 1$ is the quadrature variance of each mode in shot-noise units and $C_q = 2 \sqrt{n_q \left(n_q + 1\right)}$ quantifies the inter-mode correlations. The received signal mode is subject to the same lossy thermal channel described in~(\ref{Eq:Quantum_Channel_Received_Signal}), and the corresponding expressions~(\ref{Eq:Quantum_Channel_Received_Photons}) and (\ref{Eq:Quantum_Channel_SINR}) also apply. Target detection is performed using the correlation operator
		\begin{equation}
			\hat{c}^{[k]} = \hat{x}_r^{[k]} \hat{x}_i^{[k]} - \hat{p}_r^{[k]} \hat{p}_i^{[k]},
		\end{equation}
		where the subscript $r$ denotes the received mode~\cite{sorelli2021detecting}. Under $\mathcal{H}_1$, the expectation of the first term is
		\begin{equation}
			\begin{split}
				\left \langle \hat{x}_r^{[k]} \hat{x}_i^{[k]} \right \rangle &= \left \langle \hat{a}_t^{[k]} \hat{a}_i^{[k]} \right \rangle + \left \langle \hspace{-0.5mm} \left( \hat{a}_t^{[k]} \right)^{\hspace{-0.5mm} \dagger} \left( \hat{a}_i^{[k]} \right)^{\hspace{-0.5mm} \dagger} \right \rangle \\[1mm]
				&= 2 \sqrt{\eta^s n_q \left(n_q + 1\right)},
			\end{split}
		\end{equation}
		where~(\ref{Eq:Quantum_Channel_Received_Signal}) and~(\ref{Eq:x_quadrature}) were used. Terms involving $\left \langle \hat{a}_n^{[k]} \hat{a}_i^{[k]} \right \rangle$ vanish due to the zero-mean nature of the background noise and its statistical independence from the idler mode. Similarly, the second term expectation evaluates to
		\begin{equation}
			\left \langle \hat{p}_r^{[k]} \hat{p}_i^{[k]} \right \rangle = -2 \sqrt{\eta^s n_q \left(n_q + 1\right)},
		\end{equation}
		which leads to the expected correlation value under $\mathcal{H}_1$
		\begin{equation}
			\left \langle \hat{c}^{[k]} \right \rangle = 4 \sqrt{\eta^s n_q \left(n_q + 1\right)}.
		\end{equation}
		To compute the variance, we first evaluate the second moment
		\begin{equation}
			\begin{split}
				\hspace{-2.5mm} \left \langle \left( \hat{c}^{[k]} \right)^2 \right \rangle &= \left \langle \left( \hat{x}_r^{[k]} \hat{x}_i^{[k]} \right)^2 \right \rangle  + \left \langle \left( \hat{p}_r^{[k]} \hat{p}_i^{[k]} \right)^2 \right \rangle \\[1mm]
				& \hspace{4mm} - \left \langle \hat{x}_r^{[k]} \hat{x}_i^{[k]} \hat{p}_r^{[k]} \hat{p}_i^{[k]}  	\right \rangle - \left \langle \hat{p}_r^{[k]} \hat{p}_i^{[k]} \hat{x}_r^{[k]} \hat{x}_i^{[k]}  \right \rangle \\[1mm]
				&= 4 \left[ n_q \left( 8 \eta^s n_q + 7 \eta^s + 2 n_{w} + 1 \right) + n_{w} + 1 \right], \hspace{-3mm}
			\end{split}
		\end{equation}
		where the Wick's theorem~\cite{wick1950evaluation} was applied to factor higher-order moments. Hence, the variance of the correlation operator under $\mathcal{H}_1$ becomes
		\begin{equation}
			\sigma_{\hat{c}^{[k]}}^2 = 4 \left[ n_q \left( 4 \eta^s n_q + 3 \eta^s + 2 n_{w} + 1 \right) + n_{w} + 1 \right].
		\end{equation}
		Under $\mathcal{H}_0$, the correlation variance reduces to the same expression with $\eta^s = 0$. Finally, using~(\ref{Eq:Quantum_Channel_SINR}), the auxiliary radar parameters follow directly from the mean and variance expressions above.
\bibliographystyle{IEEEtran}
\bibliography{IEEEabrv, References}
\end{document}